\documentclass[12pt]{article}
\catcode`\@=11

\input epsf

\@addtoreset{equation}{section}
\def\theequation{\arabic{equation}}
\def\theequation{\thesection\arabic{equation}}

\newcommand{\be}{\begin{equation}}
\newcommand{\ee}{\end{equation}}
\newcommand{\ba}{\begin{eqnarray}}
\newcommand{\ea}{\end{eqnarray}}

\def\part{\partial}

\def\@normalsize{\@setsize\normalsize{15pt}\xiipt\@xiipt
\abovedisplayskip 14pt plus3pt minus3pt%
\belowdisplayskip \abovedisplayskip
\abovedisplayshortskip  \z@ plus3pt%
\belowdisplayshortskip  7pt plus3.5pt minus0pt}
\def\small{\@setsize\small{13.6pt}\xipt\@xipt
\abovedisplayskip 13pt plus3pt minus3pt%
\belowdisplayskip \abovedisplayskip
\abovedisplayshortskip  \z@ plus3pt%
\belowdisplayshortskip  7pt plus3.5pt minus0pt
\def\@listi{\parsep 4.5pt plus 2pt minus 1pt
            \itemsep \parsep
            \topsep 9pt plus 3pt minus 3pt}}

\def\underline#1{\relax\ifmmode\@@underline#1\else
        $\@@underline{\hbox{#1}}$\relax\fi}
\@twosidetrue
\relax

\catcode`@=12

\catcode`\@=11

\def\section{\@startsection{section}{1}{\z@}{3.5ex plus 1ex minus
   .2ex}{2.3ex plus .2ex}{\large\bf}}
\def\thesection{\arabic{section}.}
\def\thesubsection{\arabic{section}.\arabic{subsection}}

\def\ps@headings{\def\@oddfoot{}\def\@evenfoot{}
\def\@oddhead{\hbox{}\hfill
        \makebox[.5\textwidth]{\raggedright\ignorespaces --\thepage{}--
        \hfill }}
\def\@evenhead{\@oddhead}
\def\subsectionmark##1{\markboth{##1}{}} }
\renewcommand{\subsection}[1]{\addtocounter{subsection}{1}
\vspace{2.5mm}\par\noindent {\em \thesubsection . #1}\par
 \vspace{0.5mm} }
\ps@headings

\catcode`\@=12

\relax

\def\figcap{\section*{Figure Captions\markboth
        {FIGURECAPTIONS}{FIGURECAPTIONS}}\list
        {Fig. \arabic{enumi}:\hfill}{\settowidth\labelwidth{Fig. 999:}
        \leftmargin\labelwidth
        \advance\leftmargin\labelsep\usecounter{enumi}}}
 \relax
\def\tablecap{\section*{Table Captions\markboth
        {TABLECAPTIONS}{TABLECAPTIONS}}\list
        {Table \arabic{enumi}:\hfill}{\settowidth\labelwidth{Table 999:}
        \leftmargin\labelwidth
        \advance\leftmargin\labelsep\usecounter{enumi}}}
 \relax
\def\reflist{\section*{References\markboth
        {REFLIST}{REFLIST}}\list
        {[\arabic{enumi}]\hfill}{\settowidth\labelwidth{[999]}
        \leftmargin\labelwidth
        \advance\leftmargin\labelsep\usecounter{enumi}}}
 \relax

\catcode`\@=11

\def\marginnote#1{}
\newcount\hour
\newcount\minute
\newtoks\amorpm
\hour=\time\divide\hour by60
\minute=\time{\multiply\hour by60 \global\advance\minute by-
\hour}
\edef\standardtime{{\ifnum\hour<12 \global\amorpm={am}%
    \else\global\amorpm={pm}\advance\hour by-12 \fi
    \ifnum\hour=0 \hour=12 \fi
    \number\hour:\ifnum\minute<100\fi\number\minute\the\amorpm}}
\edef\militarytime{\number\hour:\ifnum\minute<100\fi\number\minute}
\def\draftlabel#1{{\@bsphack\if@filesw {\let\thepage\relax
  \xdef\@gtempa{\write\@auxout{\string
    \newlabel{#1}{{\@currentlabel}{\thepage}}}}}\@gtempa
    \if@nobreak \ifvmode\nobreak\fi\fi\fi\@esphack}
     \gdef\@eqnlabel{#1}}
\def\@eqnlabel{}
\def\@vacuum{}
\def\draftmarginnote#1{\marginpar{\raggedright\scriptsize\tt#1}}
\def\draft{\oddsidemargin -.5truein
        \def\@oddfoot{\sl preliminary draft \hfil
        \rm\thepage\hfil\sl\today\quad\militarytime}
        \let\@evenfoot\@oddfoot \overfullrule 3pt
        \let\label=\draftlabel
        \let\marginnote=\draftmarginnote
   
\def\@eqnnum{(\theequation)\rlap{\kern\marginparsep\tt\@eqnlabel}%
\global\let\@eqnlabel\@vacuum}  }
\def\preprint{\twocolumn\sloppy\flushbottom\parindent 1em
        \leftmargini 2em\leftmarginv .5em\leftmarginvi .5em
        \oddsidemargin -.5in    \evensidemargin -.5in
        \columnsep 15mm \footheight 0pt
        \textwidth 250mmin      \topmargin  -.4in
        \headheight 12pt \topskip .4in
        \textheight 175mm
        \footskip 0pt
        
\def\@oddhead{\thepage\hfil\addtocounter{page}{1}\thepage}
        \let\@evenhead\@oddhead \def\@oddfoot{} \def\@evenfoot{}  }
\def\titlepage{\@restonecolfalse\if@twocolumn\@restonecoltrue\onecolumn
     \else \newpage \fi \thispagestyle{empty}\c@page\z@
        \def\thefootnote{\fnsymbol{footnote}} }
\def\endtitlepage{\if@restonecol\twocolumn \else  \fi
        \def\thefootnote{\arabic{footnote}}
        \setcounter{footnote}{0}}  
\catcode`@=12
\relax

\def\ps@headings{\def\@oddfoot{}\def\@evenfoot{}
\def\@oddhead{\hbox{}\hfill
        \makebox[.5\textwidth]{\raggedright\ignorespaces --\ \thepage{}\ --
        \hfill }}
\def\@evenhead{\@oddhead}
\def\subsectionmark##1{\markboth{##1}{}} }

\ps@headings

\relax

\def\firstpage#1#2#3#4#5#6{
\begin{titlepage}
\nopagebreak
\title{\begin{flushright}
      \vspace*{-0.4in}
        {\normalsize LPT-ORSAY 01/56}\\[-6mm]
        {\normalsize ROM2F-01/18}\\[-6mm]
        {\normalsize hep-th/0107081}\\
\end{flushright}
{#3}}
\vskip .3cm
\author{ #4 \\[0.3cm] #5}
\maketitle
\nopagebreak 
\begin{abstract} {\noindent #6}
\end{abstract}
\vfill
\begin{flushleft}
\rule{16.1cm}{0.2mm}\\ 
$^{\star}${\small Unit{\'e} mixte du CNRS, UMR 8627.}\\
\today
\end{flushleft}
\end{titlepage}}
\begin{document}
\date{}
\firstpage{3118}{IC/95/34} {\large\bf Charged and Uncharged D-branes 
 in various String Theories}  
{E. Dudas$^{\,a}$, J. Mourad$^{\,a,b}$ and 
A. Sagnotti$^{\,a,c}$} 
{\small\sl
\small\sl $^a$  LPT$^\star$, B{\^a}timent 210,  Universit\'e de 
Paris-Sud\\[-5mm] \small\sl F-91405 Orsay \ FRANCE\\[-1mm]
\small\sl $^b$  LPTM, Universit\'e de 
Cergy-Pontoise, Site de Neuville III\\[-5mm] \small\sl 
F-95031 Neuville sur Oise \ FRANCE\\[-1mm]
\small\sl$^{c}$ Dipartimento di Fisica, Universit\`a di 
Roma ``Tor Vergata''\\[-5mm]
\small \sl INFN, Sezione di Roma II\\[-5mm]
\small \sl 
Via della Ricerca Scientifica 1\\[-5mm] \small\sl I-00133 Roma \ ITALY} 
{We describe how the D-brane spectra of the various
ten-dimensional string theories can be related to general properties of 
the open-closed duality, encoded in the $S$ and $P$ matrices of the
conformal field theory. We also complete the classification and 
the description of non-BPS branes in these string
theories, elucidating their non-Abelian structures and the 
nature of the corresponding super-Higgs mechanisms. We find that the
type 0 theories and their orientifolds have two types of uncharged branes,
distinguished by their couplings to the closed string tachyon.
We also find that the 0A orientifold has the unusual feature of
having charged and uncharged branes with identical world-volume 
dimensions. We conclude with some comments on fractional branes, 
elucidating their role in connection with the boundary 
states of $D_{odd}$ SU(2) WZW models.
}
\vfill\eject
\section{Introduction}

The last few years have witnessed a widespread interest in the role
of D-branes \cite{dbranes} in String Theory, after Polchinski \cite{polchinski}
elucidated their role in relation to R-R charges. To some extent, this 
interest was also spurred
by the relative simplicity with which the low-energy 
spectra of these complicated
systems can be described, as compared to those of other types of branes
responsible for non-perturbative aspects of String Theory. D-branes
and orientifold planes are also key ingredients in the construction 
of open descendants, or orientifolds \cite{descendants}, vacuum
configurations for type I strings \cite{greenschwarz} 
or for their non-supersymmetric counterparts \cite{bianchias,as95} related
to the type 0 strings \cite{dhsw}.

The study of orientifolds has so far relied on two apparently 
distinct approaches. The first, 
more rooted in the world-sheet boundary Conformal field Theory (CFT), 
has led to early constructions of orientifold models in various dimensions 
with a number of exotic features \cite{descendants}, but has not 
been widely applied to the study of D-brane configurations as such. 
The second, essentially based on the systematic use of boundary states
\cite{cs,bstates} and more rooted in the space-time picture of branes, 
has formed the basis of most D-brane studies~\cite{dbranemore,lerda}. 

Most of the early work dealt with BPS brane
configurations, or with corresponding orientifold models
with a number of residual
supersymmetries, but more recently Sen \cite{sen} initiated a systematic study 
of additional types of branes, that do not saturate the BPS bound but can
at times be stable nonetheless. Supersymmetry is in 
general fully broken by the presence of these defects, as is also 
the case when BPS branes and anti-branes are simultaneously
present, and indeed these non-BPS branes can be related to suitable
orbifolds of combinations of this type. One of the purposes of this paper is to
show how boundary CFT methods neatly encode the known properties of
D-branes, allowing a systematic discussion of additional, less
known or new, properties of charged and uncharged branes in the 
ten-dimensional non-supersymmetric orientifolds.

In orientifolds, supersymmetry breaking can be
dealt with to a level of generality comparable to what previously
attained for oriented closed strings \cite{ssclosed}, and
can be realized in essentially four different contexts. In the 
first, supersymmetry is broken from the
start, so that no gravitinos are present in the spectrum.
All models of this type are descendants of the ten-dimensional type 0 strings
of \cite{dhsw}, as in \cite{bianchias}, or of their compactifications, and
are in general fraught with tachyons. 
A special projection, however, leads to the so-called $0'$B
model, that is free of tachyons both in the open and in the closed
spectrum \cite{as95}, an interesting
property shared by corresponding orbifolds \cite{carlo}. 
In the second, the Scherk-Schwarz deformations \cite{ss} of momenta
or windings in ``parent'' oriented closed models induce supersymmetry
breaking at the {\it compactification scale} in the descendants 
\cite{dienes,ssopen}, in a way that reflects the geometry of the corresponding brane
configurations \cite{ssopen}. In the third, one resorts to a new
option provided by open strings that, consistently with
conformal invariance, can be exactly deformed by constant magnetic fields 
in internal tori \cite{ft}. As a result, supersymmetry, unbroken 
to lowest order in the closed sector, is broken in these models by 
the magnetic moment couplings of the brane excitations \cite{bachas}. 
The resulting scale of supersymmetry breaking is again tied to the 
{\it compactification scale}, and more precisely to the areas of the tori 
affected by the magnetic flux, but again one has generally to face the 
presence of tachyons \cite{luest}. T-dual descriptions 
relate this setting to configurations with branes at angles
\cite{douglas}, and special choices, corresponding to instantons in the 
internal space, can actually lead to additional supersymmetric vacua where D5 
branes, blown up uniformly on the internal tori, are exactly
described in terms of magnetized D9 branes \cite{aads}.
Finally, in the fourth \cite{sugimoto,bsb}
suitable configurations of BPS (anti)branes and orientifold planes
induce the breaking of supersymmetry at the {\it string scale} in the 
open sector. This ``brane supersymmetry breaking'', induced by non-BPS
collections of BPS objects, differs from constructions based
on stacks of genuine non-BPS branes \cite{sen}, and leads
to tachyon-free vacua, while its low-energy description \cite{dm3} 
can be related to a non-linear realization of local 
supersymmetry {\it \`a la} Volkov-Akulov, along the lines 
of \cite{samwess}.
 
The spectrum of charged branes for the $0'$B model, recently studied 
in \cite{dm1}, gives a rationale for the
results of \cite{as95,carlo}, since it displays the potential 
ingredients of orbifolds of the ten-dimensional model,  
showing in particular which combinations do not introduce tachyon 
instabilities.  Fairly enough, a full
catalogue of the available branes substantially exhibits the whole
variety of phenomena bound to be met in their presence, and
with this in mind we shall
study in detail the uncharged branes of type 0 models, the analogues of the
non-BPS branes present in supersymmetric strings. Our general conclusion
will be that, both in supersymmetric and in non-supersymmetric strings, 
tachyon instabilities are generally present in stacks of
coincident {\it uncharged} branes, that indeed do not experience a R-R
repulsion, while they are systematically absent
in stacks of coincident {\it charged} branes of a given type.
We also find that the 0A orientifold of \cite{bianchias} has {\it both} 
charged and uncharged branes with identical dimensions.  
In this paper brane configurations without
open-string tachyons will be loosely called ``stable'' although,
strictly speaking, our analysis does not
suffice to establish their stability at the quantum level.
In a similar fashion, for the sake of brevity we shall refer
to the Chan-Paton groups of brane stacks as gauge groups, even
for D0 and D$(-1)$ branes.
 
The description of branes in flat space in the formalism of
\cite{bianchias} allows a simple and general analysis
of the non-Abelian structure of their
excitations. In addition, and more importantly, the resulting
constructions do not differ
substantially from what is needed to discuss branes in (rational) curved
geometries. These settings, although conceptually richer, differ only in 
the choice of boundary CFT \cite{reckschom}, and 
no essential novelties are met in the construction of string partition
functions, although a number of important issues related to their 
space-time interpretation still await a proper clarification.

O-planes can be discussed along similar lines. Aside from the
well-known four types present in type II and type I models, 
others appear in 0A and 0B orientifolds. 
Thus, the tachyon-free 0'B orientifold \cite{as95} contains tensionless 
O9 planes with negative R-R charge, but since its R-R spectrum comprises 
all even-dimensional forms, additional ones, with $p=1,3,5,7$, 
will appear in suitable compactifications.
In a similar fashion, the other 0B orientifold in \cite{bianchias}
contains O9 planes coupling only to the dilaton, while
the third 0B orientifold has O9 planes coupling only to the closed
tachyon. The 0A orientifold \cite{bianchias}
contains uncharged O9-planes 
of four-types, depending on all possible signs of their dilaton and
tachyon couplings, and corresponding lower-dimensional O-planes 
will appear in its orbifold compactifications. 

The plan of this paper is as follows.  In Section 2 we describe the general 
rules underlying the brane spectra of the
ten-dimensional string models, and illustrate them referring to a few simple 
examples. These suffice, in particular, to 
exhibit the key role played by the $S$ and $P$ matrices in determining the 
type I D-branes. In Section 3 we study the properties of
stacks of uncharged D-branes in type 0 theories, generalizing the analysis 
in \cite{bianchias}, and show that they are
of two types, here called ${\rm D}p_{\pm}$, distinguished by the sign of their 
couplings to the closed tachyon. In Section 4 we study the properties
of stacks of non-BPS D$p$-branes in type I strings. We show that the
resulting gauge groups are orthogonal or symplectic for even $p$ and 
unitary for odd $p$, and study the cancellation of
gauge and gravitational anomalies for non-BPS D3 and D7 branes, whose massless
spectra are chiral. Section 5 is devoted to some comments on the super-Higgs
mechanism, and in particular to the issue of the gravitino mass, both  
for non-BPS branes and for non-supersymmetric configurations of
BPS (anti)branes with ``brane supersymmetry breaking''. Our conclusion
will be that, while the former can host a
standard super-Higgs mechanism, the latter are bound to lead to
non-standard realizations, along the lines
of \cite{dm3}. In Sections 6 and 7 we describe the charged and uncharged
branes of the 0B and 0A orientifolds. In all cases we determine gauge
groups and matter spectra for brane stacks and describe the
cancellation of all potential
anomalies and the resulting Wess-Zumino terms. In one of the 0B
orientifolds, we find two types of branes of identical dimensionalities,
with orthogonal and symplectic gauge
groups, respectively. In addition, for the 0A orientifold we find the novel 
feature that two types of branes with identical dimensions, one charged
and one uncharged, are simultaneously present. 
We conclude in Section 8 with some comments
on the fractional branes
of these models. These are typically characterized
by new types of R-R charges related to orbifold fixed points, 
that play an important role in orientifold
models, being directly responsible for their generalized
Green-Schwarz mechanisms \cite{ggs}. In particular, we present an 
interesting four-dimensional example with fractional D3 branes 
at a $Z_2$ orbifold singularity in the 
0'B model, and relate the peculiar features of boundaries in 
$D_{odd}$ WZW models, first noticed in \cite{completeness}, 
to the appearance of corresponding fractional branes.
Finally, the Appendix describes in some detail the compact 
notation of \cite{bianchias} used for the amplitudes. 
    
\section{General rules and some examples}

A rational {\it boundary CFT} is characterized by a  central charge
$c$ and by a finite number of characters 
$\{\chi_i\}$, of conformal weights $h_i$, acted upon by two matrices, 
$S$ and $P$. The $S$ matrix, that we shall assume symmetric and
unitary, implements on the $\{\chi_i\}$ the transformation
$\tau\to -1/\tau$, and is quite familiar from the bulk
CFT of oriented closed strings.  On the other hand, the somewhat
less familiar $P$ matrix plays a ubiquitous role in 
the determination of non-orientable spectra. It is fair to
say that, in these constructions, $P$ 
replaces the more familiar $T$ matrix, 
that implements on the $\{\chi_i\}$ the transformation $\tau\to \tau+1$ and 
acts diagonally on them as
$T_{ij} = \exp[2i \pi(h_i - c/24)] \delta_{ij}$. 
$S$ and $T$ actually determine $P$ as
\be
P = T^{1/2} S T^2 S T^{1/2} \ ,
\label{i11}
\ee
where the $T^{1/2}$ factors, with  
$T^{1/2}_{ij} = \exp[i \pi (h_i - c/24)] \delta_{ij}$, are introduced by
phase redefinitions to a convenient real basis of ``hatted'' characters for 
the M\"obius amplitude. In addition, $S^2=P^2=C$, where $C$ is the
conjugation matrix of the CFT. $S$ relates the direct and transverse
channels of the Klein-bottle and annulus amplitudes, that in the
following will be denoted by ${\cal K}$, $\tilde{\cal K}$ and ${\cal A}$, $\tilde{\cal A}$, while $P$ plays a similar 
role for the M\"obius amplitudes ${\cal M}$ and $\tilde{\cal M}$. 
Annulus and M\"obius strip bring about an additional subtlety:
their transverse channels describe 
the tree-level propagation of the {\it closed} spectrum, that lives in 
the embedding space-time, between branes that are generally
lower-dimensional, while their direct channels are one-loop amplitudes for the 
open spectra of brane modes \cite{cs}. 
As a result, in describing lower-dimensional branes one is to resort to 
character bases adapted to their reduced symmetry. 
This subtlety is particularly relevant for even-$p$ D$p$ branes, that 
have odd dimensional world volumes and therefore non-chiral spinors.

In this paper,
our focus will be on flat-space branes, and therefore the CFT's we shall
need are related to level-one orthogonal affine algebras. 
The corresponding $S$ and $P$
matrices are collected in the Appendix, together with some 
other useful properties. As we shall see, $P$
encodes the more peculiar properties of the brane spectra of orientifold
models. In order to illustrate its role, let us begin by recovering, in 
this language, a few
simple and well-known results on D-branes in type II and type I
models.

The BPS branes of the type IIB model have even-dimensional world volumes
and chiral massless spectra. In the notation of \cite{bianchias},
briefly reviewed in the Appendix, the 
annulus amplitude for a stack of these D$p$-branes is
\be
{\cal A}_{pp} = d \bar{d} \ ( V_{p-1} O_{9-p} +  O_{p-1} V_{9-p} 
-  S_{p-1} S_{9-p} -  C_{p-1} C_{9-p} ) \ , 
\label{i12}
\ee
where we have decomposed the O(8) characters with respect to the $(p-1)$
light-cone directions longitudinal to the branes. In space-time
language, $V_{p-1} O_{9-p}$ describes gauge bosons, $O_{p-1} V_{9-p}$
describes scalars (internal components of ten-dimensional vector fields) and 
$S_{p-1} S_{9-p}$ and $C_{p-1} C_{9-p}$ describe space-time fermions. As in
\cite{bianchias}, the ``complex multiplicities'' $d$($\bar{d}$) 
label the fundamental (conjugate fundamental) 
representations of the corresponding ${\rm U}(d)$ gauge groups, and
the coefficients in the 
corresponding transverse-channel amplitudes, 
\be
\tilde{\cal A}_{pp} = 2^{-(p+1)/2} d \bar{d} \ ( V_{p-1} O_{9-p} +  O_{p-1} V_{9-p} 
-  S_{p-1} S_{9-p} -  C_{p-1} C_{9-p} ) \ , 
\label{i13}
\ee
that determine both the brane tensions and their R-R charges, depend in this
case on the {\it squared absolute values} of the complex multiplicities
$d$. In the closed (transverse-channel) amplitude
(\ref{i13}), the characters have a different interpretation:
the NS-NS terms, $V_{p-1} O_{9-p}$ $+$ $O_{p-1} V_{9-p}$, describe the 
tree-level exchange of 
dilaton and internal graviton modes, while the R-R terms, $S_{p-1}
S_{9-p} + C_{p-1} C_{9-p}$, reflect the R-R charges of the various
branes. The decomposition of closed-channel contributions with respect 
to ${\rm SO}(p-1) \times {\rm SO}(9-p)$
characters plays an important role in the D9-${\rm D}p$ amplitudes,
that in the transverse channel read
\ba
\tilde{\cal A}_{9p} &=& 2^{-5}\Big[ (n \bar{d} +  \bar{n} d) \, 
( V_{p-1} O_{9-p} -  O_{p-1} V_{9-p}) \nonumber \\
&& +  
(e^{-i(p-1)\pi/4} \, n \bar{d} +  e^{i(p-1)\pi/4} \, \bar{n} d) 
(S_{p-1} S_{9-p} -  C_{p-1} C_{9-p}) \Big]
\, .
\label{i013}
\ea 
The $S$ matrix of eq. (\ref{a2}) determines the corresponding 
direct-channel amplitudes,
\ba
{\cal A}_{9p} &=& \frac{1}{2}\, (n \bar{d} +  \bar{n} d) \, 
\Big[ (O_{p-1} + V_{p-1}) (S_{9-p} + C_{9-p})-  
(S_{p-1} + C_{p-1}) (O_{9-p} + V_{9-p}) \Big]
 \nonumber \\
&& +  \frac{1}{2}
(n \bar{d} +  e^{i(p-5)\pi/2} \, \bar{n} d)(O_{p-1} - V_{p-1}) 
(S_{9-p} - C_{9-p}) \nonumber \\
&& + \frac{1}{2}
(e^{-i(p-1)\pi/2} \, n \bar{d} +  \bar{n} d)(O_{p-1} - V_{p-1}) 
(S_{9-p} - C_{9-p})  \, ,
\label{i0133}
\ea 
only consistent for odd values for $p$, that in these cases give
the chiral spectra
\be
{\cal A}_{9p} = (n \bar{d} +  \bar{n} d)( O_{p-1} S_{9-p}  +
V_{p-1} C_{9-p} - C_{p-1} O_{9-p} - S_{p-1} V_{9-p} )
\ee
for $p=1,5$ and
\ba
{\cal A}_{9p} &=& n \bar{d}\, ( O_{p-1} S_{9-p}  +
V_{p-1} C_{9-p} - S_{p-1} O_{9-p}  -
C_{p-1} V_{9-p}) \nonumber \\
&& +  \bar{n} d \, ( O_{p-1} C_{9-p}  +
V_{p-1} S_{9-p} - C_{p-1} O_{9-p}  -
S_{p-1} V_{9-p})
\ea
for $p=-1,3,7$.
The BPS 
branes for the type IIA string are very similar, being related to these
by T-dualities along odd numbers of coordinates
\cite{dbranes,polchinski}, but have odd-dimensional
world volumes, and thus non-chiral spectra.

In moving to a stack of BPS D-branes in the SO(32) type I model,
one has to face the presence of background D9-branes and 
O9-planes. These are encoded in the familiar amplitudes
\be
{\cal K} = \frac{1}{2} (V_8 - S_8) \ , \quad 
{\cal A}_{99} = \frac{n^2}{2} (V_8 - S_8) \ , \quad 
{\cal M}_{9} = - \frac{n}{2} (\hat{V}_8 - \hat{S}_8) \ , 
\label{i4}
\ee
where the ``hatted'' characters are defined in the Appendix
and $n$ equals 32 on account of tadpole cancellation, and their presence has 
two important consequences. First, the ${\rm D}p$-${\rm D}p$ amplitude
is to be supplemented with additional ones accounting for the
propagation of the bulk spectrum between the probe ${\rm D}p$ and
the background D9 and O9. Moreover, the (overall real)
Chan-Paton multiplicities of
the probe branes now lead to transverse-channel coefficients that 
are {\it perfect squares}, so that the resulting closed-channel amplitudes 
are
\ba
\tilde{\cal A}_{pp} &=& \frac{2^{-(p+1)/2}\, d^2}{2} \ ( V_{p-1} O_{9-p} +
O_{p-1} V_{9-p} -  S_{p-1} S_{9-p} -  C_{p-1} C_{9-p} ) \ , \nonumber \\
\tilde{\cal A}_{p9} &=& 2^{-5} \ n\, \times d \ ( V_{p-1} O_{9-p} -
O_{p-1} V_{9-p} + S_{p-1} S_{9-p} -  C_{p-1} C_{9-p} ) \ , \nonumber \\
\tilde{\cal M}_{p} &=& - d \ ( \hat{V}_{p-1} \hat{O}_{9-p} -
\hat{O}_{p-1} \hat{V}_{9-p} +  \hat{S}_{p-1} \hat{S}_{9-p} -  
\hat{C}_{p-1} \hat{C}_{9-p} ) \ .
\label{i5}
\ea

The closed-channel M\"obius amplitude $\tilde{\cal M}_p$ and the
D9-D$p$ amplitude $\tilde{\cal A}_{p9}$ thus
involve, again, a relative sign
between the different contributions that breaks the SO(8) space-time
symmetry. In all cases, this sign reflects the presence of $p$-dimensional
extended objects in the embedding ten-dimensional space-time, and can be 
neatly ascribed,
in the open-string channel of the M\"obius amplitude, to the additional 
parity operation
carried along by  the $\Omega$ projection when acting in the
Dirichlet-Dirichlet sector. In the D$p$-D9 sector, as we have seen,
a similar relative sign reflects the presence 
of $(9-p)$ Neumann-Dirichlet coordinates, and in more general boundary CFT's
all this is precisely in the spirit of \cite{fss}, where 
boundaries preserving only part of the bulk symmetry were studied
in detail as the proper general setting for D-brane configurations.

The breaking of the SO(8) symmetry has a 
clearcut role in the low-energy effective field theory, where the
background D9 and probe D$p$ branes would interact with the ten-dimensional
dilaton $\phi_{10}$ according to
\be
{\cal L} = - T_9 \int d^{10} x \sqrt{-g} \ e^{-\phi_{10}} - T_p 
\int d^{p+1} x \sqrt{-g} \ e^{-\phi_{10}} \ . \label{i05}
\ee 
After a reduction to $p+1$ dimensions in a compact internal volume $V$,
in terms of the $(p+1)$-dimensional dilaton, defined by
\be
e^{-\phi_{p+1}} = \sqrt{V} \ e^{-\phi_{10}} \ , \label{i06}
\ee
the resulting couplings would be proportional to
\be
{\rm D9} \ : \ \sqrt{V} \ e^{-\phi_{p+1}} \ , \qquad 
{\rm D}p \ : \ {1 \over \sqrt{V}} \ e^{-\phi_{p+1}}  \ . \label{i07}
\ee
In the closed channel, the coefficient of $V_{p-1} O_{9-p}$ thus 
determines the 
identical couplings of D9 and D$p$ branes to the fluctuation 
$\delta \phi_{p+1}$ of the
$(p+1)$-dimensional dilaton, 
while the coefficient of $O_{p-1} V_{9-p}$ determines their opposite
couplings to the ``breathing mode'', the
fluctuation $\delta V$ of the $(9-p)$-dimensional 
volume field $V$.

The direct annulus amplitudes derived from (\ref{i5}) are
\ba
&&{\cal A}_{pp} = {d^2 \over 2} \ ( V_{p-1} O_{9-p} +
O_{p-1} V_{9-p} -  S_{p-1} S_{9-p} -  C_{p-1} C_{9-p} ) \ , \label{i070} \\
&&{\cal A}_{p9} = {n\, \times d \over 2} \Bigl[
( O_{p-1}+V_{p-1}) (S_{9-p}+C_{9-p}) -( S_{p-1}+C_{p-1})
(O_{9-p}+V_{9-p}) \nonumber \\
&&+ e^{-{(9-p) i \pi \over 4}} ( O_{p-1}-V_{p-1}) (S_{9-p}-C_{9-p}) 
+ e^{-{(p-1) i \pi \over 4}} ( S_{p-1}-C_{p-1}) (O_{9-p}-V_{9-p}) \Bigr] \ ,
\nonumber  
\ea
and the D9-D$p$ amplitudes are thus inconsistent unless $p=1,5,9$, that
identify the allowed BPS branes in the type I string.
The sign in (\ref{i5}) has also a crucial
effect on the structure of the direct-channel amplitude ${\cal M}_p$,
determined by the $P$ transformation in eq. (\ref{a2}) to be
\ba
\!\!\!\!\!\!\!\!\!&&{\cal M}_{p} = - \frac{d}{2} \Biggl[    
\sin \frac{(p-5)\pi}{4} \, ( \hat{O}_{p-1} \hat{O}_{9-p} + \hat{V}_{p-1} 
\hat{V}_{9-p}) + 
\cos \frac{(p-5)\pi}{4} \, ( \hat{O}_{p-1} \hat{V}_{9-p} - \hat{V}_{p-1} 
\hat{O}_{9-p}) 
\nonumber \\
&&- i \sin \frac{(p-5)\pi}{4} ( \hat{C}_{p-1} \hat{S}_{9-p} - 
\hat{S}_{p-1} \hat{C}_{9-p}) 
- \cos \frac{(p-5)\pi}{4} ( \hat{S}_{p-1} \hat{S}_{9-p} - \hat{C}_{p-1}
\hat{C}_{9-p}) \Biggr]
\ . \label{i6}
\ea
This M\"obius projection of ${\cal A}$ 
is thus clearly inconsistent, unless $\sin (p-5)\pi/4$ vanishes, a
condition that simply recovers the other allowed BPS D-branes in 
the SO(32) type I string, 
that are indeed only D5 and D1. Moreover, since in these two cases the 
left-over cosines are equal to $\pm 1$, stacks of these D-branes have, as
is well known, USp and SO gauge groups respectively. 
Antibranes can be similarly discussed, reversing the signs of the
R-R contributions to the ${\tilde{\cal M}}_{p}$ and  
${\tilde{\cal A}}_{pq}$ amplitudes. In a similar fashion, one can
see that the 
non-supersymmetric USp(32) model of \cite{sugimoto} also allows
only D5 and D1 branes, albeit with non-supersymmetric spectra, and with SO and 
USp gauge groups, respectively.

Similar considerations determine the spectra of all charged and 
uncharged branes of the ten-dimensional string models, and the $P$
matrix always encodes interesting properties of the resulting 
orientifolds.

The following sections are devoted to a systematic discussion of the
ten-dimensional models, and all the results rest on the following, 
by now standard, criteria:
\begin{itemize}
\item[a. ] The open-string spectrum, described by the one-loop
amplitudes in the open channel, should be compatible with a correct 
space-time particle interpretation, and in particular with the appropriate
spin-statistics relation for bosons and fermions.
\item[b. ] After suitable modular transformations, the same amplitudes 
should describe the tree-level propagation of closed strings, in a way
consistent with the closed-string spectrum. This also fixes 
the relative tensions of the various branes.
\item[c. ] In the ten-dimensional orientifolds, one is also to account 
for the background O9-planes and D9-branes.
The $\Omega$ projection is encoded in the $P$ matrix of
the conformal field theory, while in the closed-channel M\"obius
amplitude the dimensions of the branes determine the decomposition of
the corresponding O(8) characters with respect to the ${\rm O}(p-1)$ subgroups.
This reflects the presence of $(9-p)$ Dirichlet coordinates in the D$p$
boundary states, that are folded into the conventional O9-planes.
\item[d. ] The charged (BPS-like) branes also couple
to the R-R fields of the closed sector, while the uncharged 
(non-BPS-like) ones couple only to NS-NS fields. 
\end{itemize}

Aside from the BPS D$p$ branes for odd (even) $p$, the type IIB (IIA) models
contain non-BPS branes for even (odd) $p$. 
These additional branes do not carry R-R charges, and are thus potentially
unstable \cite{sen}. They can be generated subjecting brane-antibrane
pairs, that in type IIB would be described by
\ba
\tilde{\cal A}_{pp} &=& 2^{-(p+1)/2} \left[ |m+n|^2  
(V_{p-1} O_{9-p} +  O_{p-1} V_{9-p}) -  
|m-n|^2  (S_{p-1} S_{9-p} +  C_{p-1} C_{9-p}) \right] \, \nonumber \\
{\cal A}_{pp} &=& (m \bar{m} + n \bar{n}) ( V_{p-1} O_{9-p} +  O_{p-1}
V_{9-p} - S_{p-1} S_{9-p} - C_{p-1} C_{9-p} ) \nonumber \\ 
&&+ (m \bar{n} + n \bar{m}) 
(O_{p-1} O_{9-p} +  V_{p-1} V_{9-p}- S_{p-1} C_{9-p} -  C_{p-1} S_{9-p}) 
\ , \label{i7}
\ea
to an orbifold operation that interchanges them, to be combined with a 
corresponding $Z_2$
operation in the closed spectrum. This involves the left space-time 
fermion number, and as a result turns
the original type IIB into type IIA. Hence,
one is finally relating non-BPS branes in type IIA to brane-antibrane pairs
in type IIB. 
In the open sector all this corresponds to identifying
$n$ and $m$ with a single charge multiplicity $N$, while rescaling 
the amplitudes by an overall factor $\frac{1}{2}$, so that
\ba
\tilde{\cal A}_{pp} &=& 2 \times 2^{-(p+1)/2} N {\bar N} \ 
(V_{p-1} O_{9-p} +  O_{p-1} V_{9-p}) \, \nonumber \\
{\cal A}_{pp} &=& N { \bar N} \ 
[( O_{p-1}+ V_{p-1})( O_{9-p}+ V_{9-p}) - 
 ( S_{p-1}+ C_{p-1})( S_{9-p}+ C_{9-p})] \ . \label{i8}
\ea
The low-lying open spectrum in (\ref{i8}) contains a {\it vector} boson, 
$(9-p)$ scalars, a 
{\it tachyon} and a {\it non-chiral fermion}, all in the adjoint 
representation of the ${\rm U}(N)$ gauge group. It is worth
stressing that, from the CFT viewpoint, these are simply branes associated to
non-diagonal bulk modular invariants, {\it i.e.} defined in settings that
are more general than the Cardy case \cite{cardy}. These non-BPS branes 
interact with the dilaton, with a tension
$\sqrt{2}$ times larger than that of the BPS branes, as needed for a
correct particle interpretation of their open-string states, and consistently
with their instability. It was 
conjectured by Sen that, after tachyon condensation, they
should decay into the vacuum. We shall return to this issue later, since
it poses interesting questions related to the super-Higgs effect
triggered by non-supersymmetric branes. 

One can discuss with no further difficulties systems of different branes,
although for the sake of brevity we shall mostly refrain from doing it
in the following sections. For instance, the strings 
stretching between 
$n$ ${\rm D}p$ and $d$ ${\rm D}q$ non-BPS branes, where
$p-q=0 \ {\rm mod} \ 2$ and, for definiteness, $p > q$, have
$q+1$ Neumann-Neumann (NN) coordinates, $p-q$
Neumann-Dirichlet (ND) coordinates and $9-p$ Dirichlet-Dirichlet
(DD) coordinates. The
corresponding annulus amplitudes read
\ba
{\tilde {\cal A}}_{pq} &=& 2 \times 2^{-(p+1)/2} \ (n {\bar d}+ {\bar n} d) \ 
\left ( V_{8-p+q}O_{p-q}- O_{8-p+q} V_{p-q} \right ) \ , \label{i08} \\
{\cal A}_{pq} &=& (n {\bar d}+ {\bar n} d) \ \left[
(O_{8-p+q} + V_{8-p+q}) (S_{p-q} + C_{p-q}) -
(S_{8-p+q} + C_{8-p+q}) (O_{p-q} + V_{p-q}) \right] \ . \nonumber  
\ea
In order to exhibit the resulting spectrum, the characters are to be
decomposed with respect to the ${\rm SO}(q-1)$ little group,
making use of eq. (\ref{a11}) of the Appendix, but in all cases there are
non-chiral space-time fermions in bi-fundamental representations 
of ${\rm U}(n) \times {\rm U}(d)$. In addition, this ${\rm D}p$-${\rm D}q$ spectrum 
contains tachyons for $|p-q| < 4$,
massless scalars for $|p-q|=4$ and only massive bosons for $|p-q| > 4$.
One can similarly write the ${\rm D}p$-${\rm D}q$ amplitude 
between a BPS and a non-BPS 
brane ($p-q=1 \ {\rm mod} \ 2$), and the corresponding 
annulus amplitudes, similar to the previous ones, read
\ba
{\tilde {\cal A}}_{pq} &=& \sqrt{2} \times 2^{-(p+1)/2} \  
(n {\bar d}+ {\bar n} d) \ 
\left ( V_{8-p+q}O_{p-q}- O_{8-p+q} V_{p-q} \right ) \ , \nonumber \\
{\cal A}_{pq} &=& (n {\bar d}+ {\bar n} d) \ \left[
(O_{8-p+q} + V_{8-p+q}) S'_{p-q}  -
S'_{8-p+q} (O_{p-q} + V_{p-q}) \right] \ . \label{i09}  
\ea  
The novelties in (\ref{i09}) are the $\sqrt{2}$ factor in the closed channel,
that results from the geometric average of BPS and non-BPS brane tensions, 
and the appearance of the non-chiral fermion characters (\ref{a01})
in the open channel, due to the odd number of ND coordinates.

Sen \cite{sen} actually introduced an additional selection rule
for these non-BPS branes. 
For instance, in the D9 case, if one starts with brane-antibrane
stacks in type IIB, as we have seen one is led to introduce a pair
of Chan-Paton multiplicities $m$ and $n$. These integers
are dimensions of sub-blocks of large Chan-Paton 
matrices, of size $m+n$, whose charges are split among the 
different states. In order to arrive at the non-BPS
D9 brane in type IIA, one begins by noting that type IIA can be obtained from
type IIB by the orbifold operation $(-1)^{G_L}$, where $G_L$ denotes the left
space-time fermion number, that in the open sector 
induces precisely
the interchange of branes and antibranes. This is reflected in 
the analytic dependence on the charge multiplicities, 
proportional to $(n-m)$, of the R-R boundary one-point functions in
eq.~(\ref{i7}). 
The induced operation is a symmetry of the open spectrum
only if $m=n=N$, and in this case the resulting projection breaks the
gauge group to the diagonal combination of the two original ones. This is
precisely the gauge group for a stack of non-BPS D9 branes in type IIA,
but this construction somehow leaves behind a {\it reducible} 
representation of the resulting Chan-Paton gauge group, with matrices
$\lambda_{V,S} = M \otimes 1_2$ and $\lambda_{O,C} = M \otimes 
\sigma_1$, with $M$ a more familiar $N\times N$ hermitian matrix 
associated to the adjoint of ${\rm U}(N)$. The reducible matrices enforce a
{\it selection rule}: in all non-vanishing amplitudes any boundary must
contain an even number of $\sigma_1$ factors. A related observation
is that some of the closed-string vertex
operators, when inserted in amplitudes, are to decorate boundaries with
additional powers of $1 \otimes \sigma_1$. These may be regarded as
end-points of cuts originating from these vertices, and one is to sum over
the possible decorations of this type, in analogy with corresponding
sums over cuts familiar from GSO projections of fermionic systems or
from orbifold constructions. The meaning of these additional insertions
can be understood noting that, when a closed-string vertex is moved 
toward a boundary, eventually it is to turn into open-string vertices with 
proper Chan-Paton assignments, that in this framework have acquired
additional labels. The R-R bulk fields of this type belong to the $S_8
\bar{C}_8$ sector and, when they come close to boundaries, turn into 
open-string ones of the $O_8$ sector.

We can now move on to complete our description of charged and
uncharged branes in the various ten-dimensional string theories. This,
as anticipated by this discussion, can be done in rather general and
efficient terms using the formalism of \cite{bianchias}. 
One of the results of this work is that, in order to obtain
non-Abelian gauge groups from stacks of coincident branes with 
{\it no open string tachyons}, the branes are to be charged under
some R-R fields of the theory under consideration. In addition, all the
ten-dimensional orientifold models allow charged tachyon-free
brane stacks of this type.
 
\section{The D-branes of type 0 string theories}

The two ten-dimensional type 0 theories \cite{dhsw} 
\ba
{\cal T}_{0A} &=& |O_8|^2 + |V_8|^2 +S_8 \bar{C}_8 +C_8 \bar{S}_8 \ ,
\nonumber \\
{\cal T}_{0B} &=& |O_8|^2 + |V_8|^2 +|S_8|^2 +|C_8|^2 \ , \label{O1}
\ea
contain a tachyon in their NS-NS
spectra. The 0B theory has four types of odd-$p$ D$p$ branes, 
characterized by a 
pair of R-R charges relative to its two R-R sectors, whose annulus
amplitudes 
\ba
\tilde{\cal A}_{pp} &=& \frac{2^{-(p+1)/2}}{2}\Biggl[ |n_1+n_2 +n_3
+n_4|^2  (V_{p-1} O_{9-p} +  O_{p-1} V_{9-p}) \nonumber \\
&& + \, |n_1+n_2 -n_3 -n_4|^2 (O_{p-1}O_{9-p}+V_{p-1} V_{9-p}) \nonumber \\
&& - \, |n_1-n_2 +n_3 -n_4|^2 (S_{p-1} S_{9-p}+C_{p-1} C_{9-p}) \nonumber \\
&& - \, |n_1-n_2 -n_3 +n_4|^2  (S_{p-1} C_{9-p}+C_{p-1} S_{9-p}) 
\Biggr]\ , \nonumber \\
{\cal A}_{pp} &=& (n_1 \bar{n}_1+n_2 \bar{n}_2+n_3 \bar{n}_3+n_4 
\bar{n}_4) ( O_{p-1} V_{9-p}+ V_{p-1} O_{9-p}) \nonumber \\
&&+\,
(n_1 \bar{n}_2+n_2 \bar{n}_1+n_3 \bar{n}_4+n_4 \bar{n}_3) 
 ( O_{p-1} O_{9-p}+ V_{p-1} V_{9-p}) \nonumber \\
&&- \, (n_1 \bar{n}_3+n_3 \bar{n}_1+n_2 \bar{n}_4+n_4 \bar{n}_2)
 ( S_{p-1} S_{9-p}+ C_{p-1} C_{9-p}) \nonumber \\
&&- \, (n_1 \bar{n}_4+n_4 \bar{n}_1+n_2 \bar{n}_3+n_3 \bar{n}_2)
 ( S_{p-1} C_{9-p}+ C_{p-1} S_{9-p}) \ , 
 \label{O2}
\ea
are essentially as in \cite{bianchias}, although of course 
they involve four types of {\it complex} charges, so that the resulting
gauge groups are
${\rm U}(n_1)\times{\rm U}(n_2)\times{\rm U}(n_3)\times{\rm U}(n_4)$.
Notice that the signs of the
couplings of these branes to the dilaton, the tachyon
and the two R-R sectors, determined by the $S_8$ and $C_8$ contributions
to $\tilde{\cal A}_{pp}$, are 
$(+,+,+,+)$, $(+,+,-,-)$, $(+,-,+,-)$, $(+,-,-,+)$, so that the
branes of the second and fourth types can be regarded as antibranes of
those of the first (${\rm D}p_1$) and third 
(${\rm D}p_2$) types. The uncharged (non-BPS-like) D9 branes of the 0A
model are now obtained as $D9_1$-${\overline{\rm
D9}}_1$ and $D9_2$-${\overline{\rm D9}}_2$ combinations, along the lines of what was
reviewed in the previous section. As in that simpler case, the orbifold
of the closed spectrum by $(-1)^{G_L}$ turns type 0B into
type 0A, and in the open sector interchanges branes and
antibranes. This operation is a symmetry when their numbers are equal,
and the end result is
\ba
\tilde{\cal A}_{pp} &=& 2^{-(p+1)/2} \left( |n+m|^2 ( V_{p-1} O_{9-p} +
O_{p-1} V_{9-p})  + |n-m|^2 \ ( O_{p-1} O_{9-p} +
V_{p-1} V_{9-p} )\right) \ , 
\nonumber  \\ 
{\cal A}_{pp} &=& (n {\bar n}+m {\bar m}) \ 
( O_{p-1}+ V_{p-1})( O_{9-p}+ V_{9-p}) \nonumber \\
&& - \, (n {\bar m}+{\bar n} m)
\ ( S_{p-1}+ C_{p-1})( S_{9-p}+ C_{9-p}) \ ,
\label{O3}
\ea
where the suffix $p$ anticipates the fact that 
$T$-duality connects the D9 branes of type 0A to corresponding 
lower-dimensional uncharged ${\rm D}p$ branes, present for even values
of $p$ in type 0B and for odd values of $p$ in type 0A. 
The four original gauge groups of charged type 0B D9 branes are
broken to a pair of diagonal combinations, so that the general
gauge group for a stack of these uncharged branes 
is ${\rm U(m)} \times {\rm U(n)}$.
With a reducible representation of the Chan-Paton group. the
vector ($V_8$) and tachyon ($O_8$) sectors, although both 
valued in the adjoint 
representation, would actually be distinguished by additional 
tensor factors $1$ or $\sigma_1$, and the same would be 
true for the two spinorial 
sectors, valued in corresponding bi-fundamental representations. 
Notice that these uncharged branes are actually of
two types, distinguished by the relative sign of their couplings to 
the closed-string tachyon. The branes of the first type, called in the 
following ${\rm D}p_{+}$ branes, have a positive coupling to the 
dilaton, {\it i.e.} a positive tension, 
and a positive coupling to the tachyon. On the other hand, the
branes of the second type, called in the following ${\rm D}p_{-}$ 
branes, have a positive tension and a negative coupling to the 
tachyon\footnote{This distinction, also noticed in the recent preprint
\cite{thompson}, that appeared while this paper was being typed, is
also manifest in the D9 spectra of the 0A orientifolds 
in \cite{bianchias}.}. 
These neutral branes of type 0
theories have a tension equal to that of corresponding BPS branes in 
type II theories, but larger than that of the charged type 0 branes by
a factor $\sqrt{2}$. Notice that, in analogy with other examples 
in the literature
\cite{bianchias,as95,klebanov,dm1}, the open spectrum
contains space-time fermions, although the closed sector contains
only bosons.

The strings  stretching between 
$n$ ${\rm D}p_{\pm}$ and $d$ ${\rm D}q_{\pm}$ branes, 
for definiteness with $p > q$ ($p-q = 0 \ {\rm mod} \ 2$), have
$q+1$ Neumann-Neumann (NN) coordinates, $p-q$
Neumann-Dirichlet (ND) coordinates and $9-p$ DD coordinates. The
corresponding annulus amplitudes read
\ba
{\tilde {\cal A}}_{pq} &=& 2^{-(p+1)/2} (n {\bar d}+ {\bar n} d) \ 
\left ( O_{8-p+q}O_{p-q}- V_{8-p+q}
V_{p-q} + V_{8-p+q}O_{p-q}- O_{8-p+q} V_{p-q} \right ) \ , \nonumber \\
{\cal A}_{pq} &=& (n {\bar d}+ {\bar n} d) \ (O_{8-p+q} + V_{8-p+q}) (S_{p-q} +
C_{p-q}) \ , \label{O4}  
\ea
and as a result the spectra contain tachyons for $|p-q| < 4$,
massless scalars for $|p-q|=4$ and only massive bosons for $|p-q| > 4$.
On the other hand, for a system of $n$ ${\rm D}p_{\pm}$ and 
$d$ ${\rm D}q_{\mp}$ branes, the amplitudes read
\ba
{\tilde {\cal A}}_{pq} &=& 2^{-(p+1)/2} (n {\bar d}+ 
{\bar n} d) \ \left ( -O_{8-p+q}O_{p-q}+ V_{8-p+q}
V_{p-q} + V_{8-p+q}O_{p-q}- O_{8-p+q} V_{p-q} \right ) \ , \nonumber \\
{\cal A}_{pq} &=& - (n {\bar d}+ {\bar n} d) \ (S_{8-p+q} + C_{8-p+q}) (O_{p-q} +
V_{p-q}) \ , \label{O5}  
\ea
and the corresponding open spectra contain non-chiral 
massless {\it fermions}.
 
All type II and type 0 uncharged branes are unstable, as signalled by
the presence of tachyons in their spectra. Some of these tachyons,
however, can be eliminated compactifying on suitable orbifolds \cite{sen},
if the branes are placed at fixed points.

\section{The non-BPS branes of type I strings}

There are two types of ten-dimensional type I strings: aside from
the usual supersymmetric model with an SO(32) gauge group \cite{greenschwarz}, 
there is indeed a second, non-supersymmetric model, with a USp(32) gauge group
\cite{sugimoto}. Whereas  in the first model there are
32 D9 branes and 32 conventional O$9_{+}$-planes, with {\it negative} tension
and {\it negative} R-R charge, in the second there are 
32 O$9_{-}$-planes, with {\it positive} tension and {\it positive} R-R
charge, together with 32 anti D9-branes \cite{sugimoto}. In the latter
case, local supersymmetry is non-linearly realized {\it \`a la}
Volkov-Akulov in the brane sector \cite{dm3}.
Both variants of type I strings have BPS D9, D5 and D1 
branes, as well as non-BPS branes for the 
remaining values of $p$. In this section we briefly present
their construction in the formalism of \cite{bianchias}, that allows a
straightforward generalization of previous results in
\cite{witten,lerda} to arbitrary stacks. 
In type I strings, these D-branes are immersed in the
proper D9 and O9 background, and therefore in this section
the $p$-$p$ annulus amplitude will always be accompanied by 
a M\"obius amplitude, originating from the O9-D$p$ exchange, 
and by a D9-D$p$ amplitude, describing the 
spectrum of strings stretched between the probe D$p$ branes and
the background D9 branes.
 
Stacks of $d$ non-BPS D$p$ branes for {\it even} $p$ ($p=0,2,4,6,8$)
can be obtained applying the orientifold projection to the corresponding
non-BPS branes of the parent type IIB. As they are uncharged with respect to 
the R-R fields,
the $\Omega$ projection acts diagonally on their Chan-Paton factors, and
therefore one expects orthogonal or symplectic gauge groups. 
The corresponding D$p$-D$p$ annulus amplitudes are thus
\ba
{\tilde {\cal A}}_{pp} &=& 2^{-(p+1)/2} d^2 \,  (V_{p-1} O_{9-p} +  O_{p-1}
V_{9-p}) \ ,  \label{I1} \\
{\cal {\cal A}}_{pp} &=& {d^2 \over 2} \ (O_{p-1}+V_{p-1})(O_{9-p}+V_{9-p})- 
2 \, \times \frac{d^2}{2} \, S'_{p-1} S'_{9-p} 
 \ , \nonumber  
\ea
where the non-chiral fermion characters $S'$ for 
odd space-time dimensions are defined in eq. (\ref{a01}). 
In this case the fermions cannot contribute to 
${\cal M}_p$, since they do not flow in $\tilde{\cal A}_{pp}$, and thus,
{\it a fortiori}, in $\tilde{\cal M}_p$. However, the presence
of {\it two} R-R contributions allows a clearcut interpretation of
the spectrum, in terms of two sectors, one symmetrized and one 
anti-symmetrized, consistently with the absence of a net contribution
to ${\cal M}_p$.
This is actually one more instance of a general 
phenomenon in boundary CFT, first met in WZW models in \cite{pss,completeness}: 
the contributions of states with identical charges that enter ${\cal A}$ 
with even multiplicities need not be matched by corresponding terms in 
${\cal M}_p$. Equivalently, in general ${\cal A}$ and ${\cal M}$ need only match 
modulo 2 for such diagonal terms, and whenever they do not match one is
describing one or more (symmetric+antisymmetric) pairs 
of representations of the gauge group.

The O9-D$p$ contribution is encoded in the M\"obius amplitude, whose precise
normalization is unambiguously determined by the non-BPS tension in
(\ref{I1}) and by the O9 tension. The result reads
\ba
{\tilde {\cal M}}_{p} \!\!\!&=&\!\!\! - \;\epsilon \sqrt{2} \ d \ 
({\hat V}_{p-1}{\hat O}_{9-p}- 
{\hat O}_{p-1}{\hat V}_{9-p}) , \label{I2} \\
{\cal M}_{p} \!\!\!&=& \!\!\! -\; {\epsilon \; d \over \sqrt{2}} 
\left[ \sin{(p-5) \pi \over 4}
({\hat O}_{p-1}{\hat O}_{9-p} \!+\! {\hat V}_{p-1}{\hat V}_{9-p})\!+\!
 \cos{(p-5) \pi \over 4}
({\hat O}_{p-1}{\hat V}_{9-p}\!-\! {\hat V}_{p-1}{\hat O}_{9-p})  \right] \, , 
\nonumber
\ea   
where the sign $\epsilon$ is $+1$ for the 
SO(32) string and $-1$
for the USp(32) string.  In relating the
open and closed channels, we have used again the $P$ 
transformation in eq. (\ref{a2}), that introduces crucial additional
factors of $\sqrt{2}$ in ${\cal M}_p$ for all even $p$.
In a similar fashion, the D9-D$p$ spectrum can be easily extracted from 
the annulus amplitudes
\ba
{\tilde {\cal A}}_{p9} &=& 2^{-5} \sqrt{2} \ 32 \times d \
(V_{p-1}O_{9-p}-O_{p-1}V_{9-p}) \ , \nonumber \\
{\cal A}_{p9} &=& 32 \ d \ \left[ (O_{p-1}+V_{p-1}) S'_{9-p} - 
S'_{p-1} (O_{9-p}+ V_{9-p}) \right] \ ,  \label{I3}
\ea
where, again, the non-chiral fermion characters $S'$ for odd space-time 
dimensions are defined in eq. (\ref{a01}) of the Appendix. 
For all these branes, the tension is $\sqrt{2}$ times larger than it would
be for BPS branes of the same dimension.
 
These expressions summarize the complete open spectra for the various 
non-BPS D$p$ branes ($p$ even) in the two type I strings. For the 
SO(32) model they are as follows:
\begin{itemize}
\item{{\bf D0-brane :} ${\rm SO}(d)$ Chan-Paton group, {\it tachyons} in the 
adjoint, {\it scalars} (including the position of the branes) in the symmetric
representation and {\it fermions} in the symmetric and antisymmetric
representations. The massless D0-D9 spectrum contains only 
{\it fermions} in the
$(32,d)$ of ${\rm SO}(32) \times {\rm SO}(d)$. The {\it tachyon} 
is projected out if $d$=1, and therefore a single D particle is
stable, as correctly pointed out in \cite{sen}.}
 
\item{{\bf D2-brane :} ${\rm SO}(d)$ gauge group, {\it tachyons} and
 {\it scalars} (including the 
position of the branes) in the symmetric representation and 
{\it fermions} in the symmetric and antisymmetric
representations. The massless D2-D9 spectrum contains only
{\it fermions} in the $(32,d)$ of 
${\rm SO}(32) \times {\rm SO}(d)$. The {\it tachyon} cannot be eliminated, and therefore
the D2 brane is unstable.}
  
\item{{\bf D4-brane :} ${\rm USp}(d)$ gauge group, {\it tachyons} in 
the adjoint representation,
{\it scalars} (including the position of the branes) in the antisymmetric
representation and {\it fermions} in the symmetric and antisymmetric
representations. The massless D4-D9 spectrum contains only 
{\it fermions} in the $(32,d)$ of ${\rm SO}(32) \times {\rm USp}(d)$. 
The {\it tachyon} cannot be eliminated, and therefore
the D4 brane is unstable.}
 
\item{{\bf D6-brane :} ${\rm USp}(d)$ gauge group, {\it tachyon} 
and {\it scalars} (including the position of 
the branes) in the antisymmetric representation and {\it fermions} in 
the symmetric and antisymmetric representations. The D6-D9 spectrum 
contains tachyons and massless {\it fermions} in the $(32,d)$ of 
${\rm SO}(32) \times {\rm USp}(d)$, and therefore the D6 brane is unstable.}  
 
\item{{\bf D8-brane :} The D8-D8 spectrum is similar to the D0-D0
spectrum above, and reduces to it upon dimensional reduction of all spatial 
coordinates. The D8-D9 spectrum contains 
tachyons and massless fermions in the $(32,d)$ of ${\rm SO}(32) \times 
{\rm SO}(d)$,
and therefore the D8 brane is unstable.}  
\end{itemize} 
 
The corresponding spectra for the USp(32) string can be
obtained from these
interchanging orthogonal and symplectic gauge groups, as well as the
related symmetric and antisymmetric representations for the matter modes. 
In particular, in this case a single D4 brane, rather than a single D0
brane, is stable.

Type I strings have also non-BPS $D(-1)$, $D3$ and $D7$ branes, but these
have a more peculiar structure, since for these dimensions
$\Omega$ can be defined only for type IIB
brane-antibrane pairs, and interchanges them. 
As a result, stacks of these additional branes have unitary gauge 
groups, and the corresponding annulus amplitudes are
\ba
{\tilde {\cal A}}_{pp} &=& {2^{-(p+1)/2} \over 2} \left[ (d+{\bar d})^2 
(V_{p-1} O_{9-p} +  O_{p-1} V_{9-p}) +  (d-{\bar d})^2 
(S_{p-1} S_{9-p}+C_{p-1} C_{9-p}) \right] \, \nonumber \\
{\cal A}_{pp} &=& d {\bar d} \ (O_{p-1} V_{9-p} +V_{p-1} O_{9-p}-
S_{p-1} S_{9-p}- C_{p-1} C_{9-p})  \nonumber \\
&& + \, {d^2+ {\bar d}^2 \over 2} (O_{p-1} O_{9-p} +V_{p-1} V_{9-p}
- S_{p-1} C_{9-p}- C_{p-1} S_{9-p}) \ . \label{I4}  
\ea
Notice that the R-R coupling in the closed channel
is actually unphysical, a familiar state of affairs whenever ``complex''
charges are present, in agreement with the fact that these 
non-BPS branes are
uncharged. As usual, the corresponding closed-channel
M\"obius amplitude
\be
{\tilde {\cal M}}_{p} = (d+{\bar d})
({\hat O}_{p-1} {\hat V}_{9-p} -{\hat V}_{p-1} {\hat O_{9-p}})
-  (d- {\bar d})
({\hat S}_{p-1} {\hat S}_{9-p} -{\hat C}_{p-1} {\hat C_{9-p}}) \ , 
\ee
can be obtained as a  ``geometric mean'' of the probe D$p$-D$p$ (cylinder) and 
background O9-O9 (Klein) amplitudes, while
\ba
{\cal M}_{p} &=& - \, {d + {\bar d} \over 2} \, \sin{(p-5) \pi \over 4}\,
({\hat O}_{p-1} {\hat O}_{9-p} +{\hat V}_{p-1} {\hat V_{9-p}}) \nonumber \\
&& - \, {d - {\bar d} \over 2} \, e^{i (p-5) \pi \over 4}\,
(-i) \, \sin{(p-5) \pi \over 4} \, ({\hat S}_{p-1} {\hat C}_{9-p}
- {\hat C}_{p-1} {\hat S_{9-p}})  
 \  \label{I5}  
\ea 
follows from it after a $P$ transformation.
We thus found, as anticipated, a ${\rm U}(d)$ gauge group, with $9-p$
{\it scalars} and  {\it fermions} in the adjoint 
representation, the latter obtained
dimensionally reducing a ten-dimensional Majorana-Weyl fermion to the
D$p$-brane
world-volume. For the D3 (D7 and D$(-1)$) brane there are also complex 
{\it tachyons} in (anti)symmetric representations,
{\it Weyl fermions} of positive chirality in the symmetric representation and
{\it Weyl  fermions} of negative chirality in the antisymmetric 
representation of
the gauge group. Finally, the low-lying D$p$-D9 spectrum, 
encoded in the amplitudes
\ba
{\tilde {\cal A}}_{p9} &\!=\!& 2^{-5} \, n \, \left[ (d\!+\!{\bar d}) 
(V_{p-1}O_{9-p} \!-\! O_{p-1}V_{9-p}) \!-\! i (d\!-\!{\bar d}) (S_{p-1}
S_{9-p}\!-\! C_{p-1} C_{9-p}) \right] \ , \nonumber \\
{\cal A}_{p9} &=& d \, n \ (O_{p-1} S_{9-p} +V_{p-1} C_{9-p}-
C_{p-1} O_{9-p}- S_{p-1} V_{9-p})   \nonumber \\
&& + \, {\bar d}\, n \ (O_{p-1} C_{9-p} +V_{p-1} S_{9-p}
- S_{p-1} O_{9-p}- C_{p-1} V_{9-p}) 
 \ , \label{I6}  
\ea
where $n$, equal to 32, is the D9 Chan-Paton multiplicity,
comprises in both cases massless {\it Weyl fermions} in 
the $(32,d)$ of ${\rm SO}(32) \times {\rm U}(d)$, and for the D7
brane also complex {\it tachyons} in the $(32,d)$. These chiral spectra embody a
non-trivial cancellation of irreducible gauge anomalies between
the D$p$-D$p$ and D$p$-D9 sectors. The corresponding results for
the USp(32) type I string can again be obtained interchanging symmetric and 
antisymmetric representations, while also flipping the (space-time and internal)
chiralities in the D$p$-D9 sector. Notice that the $D(-1)$ brane 
(D-instanton) in the SO(32) string and the D3 brane in the USp(32)
string are {\it stable}, being free of tachyons. The tensions
of these branes are {\it twice} the values one would expect for BPS 
type I branes of the same dimension, if they existed.

Since the non-BPS D3 and D7 branes have chiral spectra, it is instructive 
to verify how the resulting anomalies cancel.
The potentially anomalous groups
are the ${\rm SO}(1,p)$ Lorentz group relative to the world-volume 
of the branes, the
transverse ${\rm SO}(9-p)$ rotation group, the ten-dimensional SO(32) gauge group and, finally,
the ${\rm U}(d)$ gauge group for the brane stack \cite{ghm}. 
The anomaly polynomials are the 6 and 10-form contributions to
\ba
{{\hat A(R)}\, {[\hat A(N)]^{-1}}}\!\!\!\!\!\!&&\!\!\!\!\!\!
\Big[\ {{1}\over{2}}(ch_+(N)-ch_-(N)) \ tr_{ad}e^{iG} \nonumber \\
&+&  ch_-(N)\ tr_S e^{iG} \, -\,  ch_+(N)\ tr_A e^{iG}\, -\, 
 tr e^{iF} \ tr e^{iG} \, \Big] \ , 
\label{an}
\ea
where $R$, $N$, $G$ and $F$ are curvature forms for
the ${\rm SO}(1,9)$, ${\rm SO}(9-p)$, ${\rm U}(d)$ and ${\rm SO}(32)$ gauge groups.
An explicit calculation reveals that 
all  irreducible anomalies indeed cancel in (\ref{an}), while the residual
anomaly polynomials are 
\ba
I_6&=& \ Y_2 \ X_4 \ ,\\
I_{10}&=&  \ Y_2 \ X_8\, +\, Y_6 \ X_4\, - 2 \, N \ Y_2 \ Y_6\ . 
\label{ano}
\ea
Here
\be 
Y_2=-i\; tr G \ , \qquad Y_6= {{i}\over{6}}\, 
tr G^3 + i \ tr G \, {p_1 (R) \over 48} - {i \over 24} tr G \ N^2 \ ,
\ee
\ba
X_4&=& - \, p_1(R) \, - \, {1\over 2} \,tr F^2  \ , \nonumber \\
X_8&=& {{tr F^4}\over{24}}\, +\, {{p_1 (R) \; tr F^2} \over{96}}
\, +\, {{3p_1(R)^2-4p_2(R)} \over{192}} 
\ea
factorize the ten dimensional anomaly polynomial as $I_{12}=X_4 \ X_8$,
and
\be
p_1(R) = - \, \frac{1}{2} \, tr R^2 \ , \qquad p_2 (R) = 
\, \frac{1}{8} \, [ (tr R^2)^2- 2 \, tr
R^4] \ .
\ee

It should be appreciated that the adjoint fermions play a crucial role in
canceling the irreducible part of the anomaly due to the
curvature of the normal bundle. The residual anomaly polynomials in (\ref{ano})
can then be canceled by the Wess-Zumino terms
\ba
S_{WZ}(D_3)&=&T_1 \int_{D_3} Y_2 \; B_2 \ , \nonumber \\
S_{WZ}(D7)&=&T_5 \int_{D_7} {Y_2} \; B_6\, +\, T_1 \int_{D_7}
Y_6 \; B_2 
\ea
in the effective actions for the D-branes. In verifying the cancellation
one needs to use the relations
\be 
\delta(D3)|_{D3}= \chi(N) \ ,\quad  \delta(D7)|_{D7}=N \ \label{I106}
\ee
between the $\delta$ functions on the brane world-volumes and the
Euler characters of the normal bundles \cite{bott}. The mixing of ${\rm U}(1)$ gauge
fields with R-R forms in the Wess-Zumino couplings implies that they become
massive, leaving only ${\rm SU}(d)$ unbroken gauge groups.   
It is interesting to note that wrapping the D7 brane on a magnetized torus 
would give rise to a non-BPS, but nevertheless charged, D5 brane.
 
\section{Comments on the super-Higgs mechanism on non-BPS branes}

The D-branes and O-planes of supersymmetric (type II or type I)
strings can trigger the complete breaking of
supersymmetry, and this asks for an understanding of the corresponding 
super-Higgs mechanisms. This issue was recently analyzed in \cite{dm3}
for the non-BPS combinations present in the
USp(32) type I string \cite{sugimoto}, where supersymmetry 
is broken at the tree level in the open sector due to the 
simultaneous presence of
$\overline{\rm D}9$ branes and ${\rm O9}_{-}$ planes. More precisely,
the minimal ten-dimensional 
supersymmetry is realized linearly in the closed sector and non-linearly
in the open sector, and the goldstino, present in the massless brane spectrum, 
has consistent interactions, although the Majorana-Weyl 
ten-dimensional gravitino does not 
allow a mass term in the Lagrangian. The super-Higgs mechanism 
is thus taking an unconventional form in this ten-dimensional model. 
This peculiar fact is also revealed
by a simple counting: the 64 combined degrees of freedom of the 
massless gravitino and of the brane goldstino are far fewer 
than the 128 proper of a
massive ten-dimensional gravitino. Still, they are compatible 
with a massive nine-dimensional gravitino, 
that together with the corresponding internal component $\psi_9$ would have precisely
64 components, and indeed the
theory has a vacuum with an SO(1,8) symmetry group, smaller than the
maximal symmetry groups compatible with the ten-dimensional 
brane world-volume \cite{dm2}. 
Further arguments to this effect are provided in \cite{sw}. 

A similar phenomenon appears at work also in lower-dimensional
models with ``brane supersymmetry breaking'' \cite{bsb}. In all
these cases one has non-BPS combinations of BPS (anti)branes, that in
type I are D9, D5 and D1, and corresponding O-planes. The simplest
manifestation of D5 branes in this context is provided by the 
six-dimensional $T^4 /Z_2$ type I model with
32 D9 branes and 32 $\overline{\rm D}5$ branes \cite{bsb}, with 
a supersymmetric closed spectrum including 16 (1,0) tensor multiplets from
the twisted sector, one per fixed point. In world-sheet language,
the $\Omega$ projection 
has here a flipped sign in the whole twisted sector,
a feature reminiscent of the WZW models discussed in \cite{pss,completeness}.
In space-time language, the vacuum includes 32 ${\rm O}9_{+}$ planes and
32 ${\rm O}5_{-}$ planes, and as a result supersymmetry is broken at the
string scale on all $\overline{\rm D}5$ branes sitting at orbifold fixed points. 
The fermion counting relative to a D5-brane world volume
now goes as follows: a massive six-dimensional gravitino would 
have 32 degrees of freedom, 
while the original massless one has 12, and the brane goldstino only 4,
but again a similar five-dimensional counting does not 
contradict a standard lower-dimensional 
realization of the super-Higgs mechanism, since a massive
five-dimensional gravitino together with the corresponding internal component $\psi_5$
would have 16 degrees of freedom. These two examples thus capture all basic
features, in this respect, of four-dimensional 
models with ``brane supersymmetry breaking''.

A similar question is clearly raised by the other non-BPS branes 
of type II and type I theories discussed in the previous
sections. Again, the D$p$-D$p$ sector always contains a candidate 
goldstino, with the correct chirality, and a
quick case-by-case analysis shows that 
massless gravitinos and goldstinos always provide the proper numbers of
degrees of freedom for massive gravitinos on brane world-volumes. 
For instance, the non-BPS D9 branes of type IIA have in their world volumes
a Majorana goldstino, with 16 degrees of freedom, that can be eaten by
a massless ten-dimensional Majorana gravitino, with 112 degrees of freedom, to give
a massive ten-dimensional Majorana gravitino, with 128 degrees of freedom. As
another example, the D8 branes of type I have in their world volumes
a nine-dimensional Majorana goldstino, that can turn a massless
nine-dimensional Majorana gravitino, with 48
degrees of freedom, into a massive one. As a last example, the D3 branes
of type I have in their world volumes four Majorana goldstinos, which
can mix with the four Majorana gravitinos to give four massive Majorana
gravitinos. All these examples thus point toward standard realizations of the 
super-Higgs mechanism on non-BPS branes. In principle, this could be 
verified explicitly from 
the effective Lagrangian for the ten-dimensional supergravities coupled to non-BPS D$p$
branes, following lines similar to those in \cite{dm3}.
Further support to this conjecture is provided by 
the classical supergravity solution corresponding to D$p$-$\overline{\rm D}p$ systems
\cite{bmo}, that has the full
${\rm SO}(p+1)$ symmetry along directions tangential to the world-volume of
the non-BPS branes, as expected for a conventional
super-Higgs mechanism on them. 
However, the classical backgrounds of
non-BPS configurations typically have naked metric singularities, whose
resolution is clearly important in order to gain a 
better control of these super-Higgs mechanisms and of the corresponding
low-energy physics.

\section{The D-branes of the 0B orientifolds}

The 0B orientifolds were constructed in \cite{bianchias,as95}. 
There are three possible choices,
generated by $\Omega$, $\Omega \times (-1)^{G_L}$, where $G_L$ is the left 
space-time fermion number, and $\Omega \times (-1)^{F_L}$, where $F_L$ is 
the left world-sheet fermion number. The last model, 
usually called $0'$B in
the literature, is non-tachyonic and contains chiral fermions in the D9
open spectrum \cite{as95}. Non-tachyonic orbifold compactifications 
of the $0'$B orientifold were studied in \cite{carlo}, while the 
spectrum of its charged branes and their anomaly cancellation mechanisms 
were recently described in \cite{dm1}, where it was shown that
the theory contains D9, D7, D5, D3, D1 and ${\rm D}(-1)$ charged branes, 
all with individual non-tachyonic chiral spectra based on unitary 
gauge groups \footnote{Tachyonic modes, however, are present in 
D$p$-D$q$ exchange spectra, for $|p-q|<4$, as is also the case
for the type IIB string.}.

The orientifold projection $\Omega \times (-1)^{F_L}$ leading to the
$0'$B theory corresponds to the Klein bottle amplitude
\ba
{\cal K} &=& {1 \over 2} \ (-O_8+V_8-S_8+C_8) \ , \nonumber \\
{\tilde {\cal K}} &=& - 2^5 \, S_8 \ , \label{B1}
\ea 
and the tachyon, odd under the
orientifold projection, is thus removed from the spectrum, while a net
number of D9 branes is needed to compensate the R-R charge 
of the O9 planes.
As displayed in (\ref{B1}), the O-planes are in this case 
rather peculiar, since they have a vanishing tension. The D-brane
spectrum can actually be anticipated recalling that, as described in
Section 3, the parent 0B theory has two
types of charged branes, whose R-R charges 
have the overall signs $(+,+)$ and $(+,-)$, and two
corresponding types of charged antibranes. The orientifold
acts collectively on the two sets of R-R fields $(A,A')$ in (\ref{B1}) 
according to $\Omega (A,A') = (-A^T,{A'}^T)
$, and on the different R-R forms this translates into  
$\Omega (A^{(0)},A'^{(0)}) = (-A^{(0)},A'^{(0)})
$,  $\Omega (A^{(2)},A'^{(2)}) = (A^{(2)},- A'^{(2)})$ and
 $\Omega (A^{(4)},A'^{(4)}) = (-A^{(4)},A'^{(4)})$.  Hence, in all cases
only the combination $(+,+) + (-,+)$, and the
corresponding antibrane, are invariant under $\Omega$ or, in the
language of Section 3,
configurations with an arbitrary number $d$ of D$p_1$-D$p_2$ pairs, whose 
members are interchanged by the $\Omega$ projection. All this is 
reminiscent of what we saw
for the D3 and D7 branes of the type I models, and
one can anticipate the occurrence
of unitary gauge groups and of a {\it BPS-like} behavior
for these charged branes, with no mutual tree-level interaction energy.
 
Let us briefly review the explicit construction of the charged D$p$ branes
present in this model. To begin with, the D9 branes are described by the 
amplitudes \cite{as95}
\ba
\tilde{\cal A}_{99} &=& \frac{2^{-5}}{4} \left[  
(n + \bar{n})^2 (V_8 - S_8) -  (n - \bar{n})^2 
(O_8 - C_8)  \right] \ ,   \nonumber \\
\tilde{\cal M}_9 &=&  (n+\bar{n}) \, \hat{S}_8 \ , 
\nonumber \\
{\cal A}_{99} &=& n \, {\bar n} \, V_8 \ - \ {n^2 + {\bar n}^2 \over 2} 
\, S_8 \ , \nonumber \\
{\cal M}_9 &=& {n+{\bar n} \over 2} \, {\hat S}_8 \ , \label{B01} 
\ea
where $n$ is a ``complex'' Chan-Paton multiplicity, and
the R-R tadpoles require that $n={\bar n}=32$, thus fixing the U(32) 
gauge group. The massless D9 spectrum is chiral, since it includes 
Weyl fermions in the antisymmetric representation, precisely as needed
to compensate the bulk contribution to irreducible gravitational anomalies. 
Notice that $\tilde{\cal A}$ contains an unphysical tachyon coupling 
in the closed sector, proportional to $(n -
\bar{n})$, consistently with the fact that the closed-string tachyon was 
actually removed by the orientifold projection.

The D5 and D1 branes are described by 
\ba
\tilde{\cal A}_{pp} &=& \frac{2^{-(p+1)/2}}{4} \Big[  
(d + \bar{d})^2 ( V_{p-1} O_{9-p} +  O_{p-1}
V_{9-p} - S_{p-1}S_{9-p}-C_{p-1}C_{9-p}) \nonumber \\
&& - \, (d - \bar{d})^2 (O_{p-1}O_{9-p}+V_{p-1} V_{9-p} - 
S_{p-1} C_{9-p}- C_{p-1} S_{9-p})  \Big]  \ , \nonumber \\
\tilde{\cal M}_p &=&  (d+\bar{d}) (\hat{S}_{p-1} \hat{S}_{9-p} - 
\hat{C}_{p-1} \hat{C}_{9-p} ) \ ,  
\nonumber \\
{\cal A}_{pp} &=& d {\bar d} \ (V_{p-1}O_{9-p}+ O_{p-1}V_{9-p})
 - {d^2 + {\bar d}^2 \over 2} (S_{p-1}S_{9-p}+ C_{p-1}C_{9-p})  \ , \nonumber
\\
{\cal M}_p &=& \epsilon \, {d+{\bar d} \over 2}\, 
({\hat S}_{p-1}{\hat S}_{9-p}-{\hat C}_{p-1}{\hat C}_{9-p}) \ , \label{B02} 
\ea
where, as in the previous sections, the relative sign between the two 
contributions to ${\cal M}_p$ reflects the dimensionality of the branes,
while the sign $\epsilon$, $+1$ ($-1$) for D5(D1) branes, 
is dictated by the corresponding $P$ matrices. 
In both cases the gauge group is ${\rm U}(d)$, and in
both cases the massless spectra contain Weyl fermions of one chirality 
in the symmetric representation, together with
Weyl fermions of the opposite chirality in the antisymmetric
representation. In addition, the D9-D$p$ spectrum is described by
\be
{\cal A}_{p9} = (n {\bar d} + {\bar n} d ) (O_{p-1}S_{9-p}+
V_{p-1}C_{9-p})
- (n d + {\bar n} {\bar d} ) (S_{p-1}V_{9-p}+
C_{p-1}O_{9-p}) \ ,  
\ee
and therefore \cite{dm1} both the D5-D9 and
D1-D9 spectra have
chiral fermions in bi-fundamental representations and
no tachyons.

On the other hand, the D7, D$(-1)$ and D3 branes present a subtlety \cite{dm1}, 
since they couple to the 0-form and to the
4-form from the R-R sector, described by the $C_8$ character in
(\ref{B1}).
This subtlety is precisely encoded in the $P$ matrix, and indeed,
starting again from the closed channel and reverting to the open channel
by $S$ and $P$ transformations gives the consistent amplitudes
\ba
\tilde{\cal A}_{pp} &=& \frac{2^{-(p+1)/2}}{4} \Big[  
(d + \bar{d})^2 ( V_{p-1} O_{9-p} +  O_{p-1}
V_{9-p} - S_{p-1} C_{9-p} - C_{p-1} S_{9-p}) \nonumber \\
&& - (d - \bar{d})^2 (O_{p-1}O_{9-p}+V_{p-1} V_{9-p} - S_{p-1}S_{9-p}
- C_{p-1} C_{9-p})  \Big] \ , \nonumber \\
\tilde{\cal M}_p &=&  - \, (d-\bar{d}) (\hat{S}_{p-1} \hat{S}_{9-p} - 
\hat{C}_{p-1} \hat{C}_{9-p} )
\, \nonumber \\
{\cal A}_{pp} &=& d {\bar d} \, (V_{p-1}O_{9-p}+ O_{p-1}V_{9-p})
 - {d^2 + {\bar d}^2 \over 2} (S_{p-1}C_{9-p}+ C_{p-1}S_{9-p})  \ , \nonumber
\\
{\cal M}_p &=& \epsilon \, {d-{\bar d} \over 2}\,  
({\hat S}_{p-1}{\hat C}_{9-p}-{\hat C}_{p-1}{\hat S}_{9-p}) \ , \label{B03} 
\ea
where the sign $\epsilon$ is $+1$ ($-1$) for the D3 (D7 and D$(-1)$) branes.   
 
In addition, the D9-D$p$ spectra are described by
\ba
{\cal A}_{p9} &=& {\bar n} d  \, (O_{p-1}S_{9-p}+ V_{p-1}C_{9-p})
+ n {\bar d} \, (O_{p-1}C_{9-p}+ V_{p-1}S_{9-p}) \nonumber \\
&& - n d  \, (S_{p-1}O_{9-p}+ C_{p-1}V_{9-p})
- {\bar n} {\bar d} \, (S_{p-1}V_{9-p}+ C_{p-1}O_{9-p}) \ ,  \label{B033}
\ea
and therefore \cite{dm1} there are tachyonic modes in the D9-D7 case.
Massless fermions in bi-fundamental representations are present
in both the D9-D7 and D9-D3 mixed spectra. Notice that all
these charged branes, whose anomaly cancellation was studied in
\cite{dm1},  are BPS-like, {\it i.e.} there is no brane-brane
interaction to lowest order.

We can now turn to the uncharged branes present in the
$0'$B orientifold, that exist for $p=$0,2,4,6,8, and whose spectra can
again be determined starting from the uncharged branes of the parent 0B
model. As we saw in Section 3, in type 0 theories the uncharged branes are of
two types, ${\rm D}p_{+}$ and ${\rm D}p_{-}$, distinguished by the sign of
their coupling to the tachyon. In this case the orientifold projection 
interchanges them, since it removes the closed string tachyon, and
therefore the final invariant combinations contain an arbitrary number
$d$ of ${\rm D}p_{+}$-${\rm D}p_{-}$ pairs, with corresponding ${\rm U}(d)$
gauge groups. 

The consistency requirements summarized in Section 2 
uniquely determine the D$p$-D$p$ amplitude, and 
therefore the spectrum of a generic stack of these uncharged D-branes
can be read from
\ba
{\tilde A}_{pp} &=& {2^{-(p+1)/2} \over 2} \left[ (d+{\bar d})^2 
( V_{p-1} O_{9-p} +  O_{p-1}
V_{9-p}) -  (d-{\bar d})^2 (O_{p-1}O_{9-p}+V_{p-1} V_{9-p}) \right] \nonumber \ , \nonumber \\
{\cal A}_{pp} &=& d {\bar d} \ (O_{p-1}+ V_{p-1}) (O_{9-p}+ V_{9-p}) \, - \, 2 \, \times \, \frac{(d^2+ {\bar d}^2)}{2} \,   S'_{p-1} S'_{9-p}  \ ,
\label{B2}  
\ea
where the fermions are now described by the non-chiral characters 
(\ref{a01}) appropriate for these odd-dimensional world-volumes.
In contrast with the parent 0B theory, in this case there is
a single type of uncharged brane, that couples only to the dilaton.
A peculiar and interesting feature of these uncharged branes is the
lack of D$p$-O9 propagation, since only the 
R-R fields couple to the O9-planes while, on the
contrary, only the NS-NS dilaton couples to these D$p$-branes. 
As in Section 4, the lack of a M\"obius contribution
implies the presence of two (symmetric+antisymmetric) pairs of R sectors.
In addition, there are D$p$-D9 string excitations described by  
\ba
{\cal A}_{p9} &\!\!=\!\!&  \ 
(n {\bar d} + {\bar n} d)(O_{p-1} + V_{p-1}) S'_{9-p} 
-  (n d + {\bar n} {\bar d}) S'_{p-1} (O_{9-p} + V_{9-p}) \ ,  \label{B3}
 \\
{\tilde {\cal A}}_{p9} &\!=\!& {2^{-5} \over \sqrt{2} } \left[ 
(n \!+\! {\bar n})(d+{\bar d})
(V_{p-1}O_{9-p} \!-\! O_{p-1}V_{9-p}) \!-\! (n\!-\!{\bar n}) 
(d\!-\!{\bar d}) (O_{p-1}
O_{9-p} \!-\! V_{p-1} V_{9-p}) \right] \ , \nonumber 
\ea
where $n$, equal to 32, is the D9 Chan-Paton multiplicity. 

We can now turn to the orientifold of the 0B model obtained by
the standard $\Omega$ projection, whose Klein bottle amplitude reads 
\cite{bianchias}
\ba
{\cal K} &=& {1 \over 2} \ (O_8+V_8-S_8-C_8) \ , \nonumber \\
{\tilde {\cal K}} &=&  2^5 \ V_8 \ . \label{B4}
\ea  
In this case there is no consistency condition 
asking for a D9 sector, aside from a
dilaton tadpole, that just signals the need for a non-trivial 
space-time background. If, as in \cite{bianchias}, one includes it,
the resulting open spectrum is described by
\ba
{\cal A}_{99} &=& {n_1^2 + n_2^2 + n_3^3 + n_4^2 \over 2} \, V_8
+ (n_1n_2+n_3n_4)\, O_8 \nonumber \\
&& -\, (n_1n_3+n_2n_4) \, S_8
-  (n_1n_4+n_2n_3) \, C_8 \ , \nonumber \\
{\cal M}_9 &=& -\, \frac{1}{2} \, (n_1+n_2+n_3+n_4) \, \hat{V}_8
 \ , \label{B04}
\ea
and the induced R-R tadpole cancellation conditions require
that $n_1=n_2$ and $n_3=n_4$, thus determining the family
of D9 gauge groups ${\rm SO}(n_1)^2
\times {\rm SO}(n_3)^2$, while the
dilaton tadpole is canceled by the unique choice $n_1+n_3=32$.
In addition, as for type I strings, one has the option of reversing 
the M\"obius projection altogether, thus obtaining symplectic gauge groups but,
as for the USp(32) type I string, it is then impossible to cancel the 
resulting dilaton tadpole. 
Notice that, in both cases, no choice for the charges $n_1$ and 
$n_3$ can eliminate the open-string tachyons, since we are forced to add
branes and antibranes in equal numbers.

Since the projected closed spectrum contains two R-R two-forms, 
the model should also have charged D1 and D5 branes. Moreover, the two types of
branes of the parent 0B, D$p_1$ and D$p_2$, separately invariant
under $\Omega$, will both appear in this orientifold, together with
their antibranes, albeit with projected gauge groups. 
The resulting D$p$-D$p$ annulus amplitudes are simply 
determined by the dimensional 
reduction of the  D9-D9 amplitude in \cite{bianchias} 
to the D$p$ world volume, and read
\ba
{\cal A}_{pp} &=& {d_1^2 + d_2^2 + d_3^3 + d_4^2 \over 2} (V_{p-1}
O_{9-p}+O_{p-1}V_{9-p}) + (d_1d_2+d_3d_4) (O_{p-1} O_{9-p}+ 
V_{p-1}V_{9-p}) \nonumber \\
&& -\, (d_1d_3+d_2d_4) (S_{p-1} S_{9-p}+ C_{p-1} C_{9-p})
-  (d_1d_4+d_2d_3) (S_{p-1} C_{9-p}+ C_{p-1} S_{9-p}) \ , \nonumber \\
{\tilde A}_{pp} &=& {2^{-(p+1)/2} \over 4} \Big[ (d_1+d_2+d_3+d_4)^2  
(V_{p-1} O_{9-p} +  O_{p-1} V_{9-p}) \nonumber \\
&& +\, (d_1+d_2-d_3-d_4)^2 (O_{p-1}O_{9-p}+V_{p-1} V_{9-p}) \Big] \nonumber \\
&& - \, {2^{-(p+1)/2} \over 4} \Big[ (d_1-d_2+d_3-d_4)^2 
(S_{p-1} S_{9-p}+C_{p-1} C_{9-p}) \nonumber \\
&& +\, (d_1-d_2-d_3+d_4)^2 (S_{p-1} C_{9-p}+
C_{p-1} S_{9-p}) \Big] \ . \label{B5}
\ea
Eq. (\ref{B5}) clearly displays the couplings of these branes to the two
sets of R-R fields. The peculiar properties of the lower-dimensional branes
are again fully encoded in the $P$ matrix and, taking into account 
the proper character decompositions,
\be
{\cal M}_p = {d_1+d_2+d_3+d_4 \over 2} \ \epsilon \ ({\hat V}_{p-1}{\hat O}_{9-p}-
{\hat O}_{p-1}{\hat V}_{9-p}) \ , \label{B6}
\ee 
where the sign $\epsilon$ is $+1$ for the D5 branes and $-1$ 
for the D1 branes.
The resulting gauge groups are thus ${\rm USp}(d_1) \times {\rm USp}(d_2) 
\times {\rm USp}(d_3)
\times {\rm USp}(d_4)$ for the D5 branes and  ${\rm SO}(d_1) \times 
{\rm SO}(d_2) \times {\rm SO}(d_3)
\times {\rm SO}(d_4)$ for the D1 branes. In a similar fashion, for the 
ten-dimensional model with symplectic gauge groups, one would find 
orthogonal groups for 
D5 branes and symplectic groups for D1 branes. In all these
cases, one can obtain {\it tachyon-free
configurations} considering only branes, {\it i.e.} setting
$d_2=d_4=0$, with gauge groups of the type
${\rm USp}(d_1) \times {\rm USp}(d_3)$ for the D5 branes and ${\rm SO}(d_1) \times {\rm SO}(d_3)$
for the D1 branes, and {\it vice versa} for the other class of 
ten-dimensional models. Moreover, equal numbers of D$p_1$ and
D$p_2$ branes give gauge groups ${\rm USp}(d)^2$ 
(${\rm SO}(d)^2$) for D5(D1) branes, and {\it vice versa} for the second
class of ten-dimensional models, and these  D$p$-D$p$ configurations are
BPS-like, {\it i.e.} there is no net tree-level brane-brane interaction.
As we have seen, this property is shared not only by
the BPS branes of type II and type I models, but also by the
D9 \cite{as95} and the other charged D$p$ 
branes \cite{dm1} of the $0'$B orientifold. 
One can similarly obtain the additional spectra related to D$p$-D9 exchanges. 
For instance, the D5-D9 spectrum is encoded in the amplitude
\ba
{\cal A}_{59} &=& (n_1 d_1 \!+\! n_2 d_2 \!+\! n_3 d_3 + n_4 d_4) \ 
(O_{4} C_4+ V_{4}
S_{4}) \nonumber \\
&&+\, (n_1 d_2 + n_2 d_1 + n_3 d_4 + n_4 d_3) \ (O_{4} S_4+ V_{4}
C_{4}) \nonumber \\ 
&&-\,  (n_1 d_4 + n_2 d_3 + n_3 d_2 + n_4 d_1) \ (C_{4} O_4+ S_{4}
V_{4}) \nonumber \\
&&- \, (n_1 d_3 + n_2 d_4 + n_3 d_1 + n_4 d_2) \ (S_{4} O_4+ C_{4}
V_{4}) \ , \label{B06} 
\ea
with massless scalars and fermions in bi-fundamental representations,
while the D1-D9 amplitude contains fermions of opposite chiralities, 
obtained interchanging the $S$ and $C$ sectors.
  
The chiral spectra of these charged branes imply 
an anomaly inflow from the D9 branes, and it 
is instructive to study the corresponding anomaly cancellation. 
The anomaly polynomial of the ten-dimensional model (\ref{B4}) is 
in this case
\be
I_{12} = X_4^{(+)} X_8^{(-)} +  X_4^{(-)} X_8^{(+)} \ , \label{B060} 
\ee
with
\ba
X_4^{(+)} &=& {1 \over 4} \, (tr F_4^2 - tr F_3^2) \ ,  \qquad
X_4^{(-)} = {1 \over 4} \, (tr F_2^2 - tr F_1^2) \ , \nonumber \\ 
X_8^{(+)} &=&  {1 \over 24} \, (tr F_3^4 - tr F_4^4)-  
{tr R^2 \over 192} \, (tr F_3^2 - tr F_4^2) \ , \nonumber \\
X_8^{(-)} &=&  {1 \over 24} \, (tr F_1^4 - tr F_2^4)-  
{tr R^2 \over 192} \, (tr F_1^2 - tr F_2^2) \ , \label{B061} 
\ea
where $F_1,  \cdots, F_4$ denote the four D9 brane gauge groups.
In order to derive unambiguously Bianchi identities and field equations
of the R-R-forms, one also needs the anomaly polynomials for the D1 and D5
branes, 
\ba
I_4 &=&  d_1 X_4^{(+)} + d_3 X_4^{(-)} \ , \label{B0061} \\
I_8 &=& {d_1 \over 2} \, X_8^{(+)} + {d_3 \over 2} \, X_8^{(-)} + 
Y_4^{(-)} ( X_4^{(-)}+ {d_1 \over 2} \, \chi
(N)) + Y_4^{(+)} ( X_4^{(+)}+ {d_3 \over 2} \, \chi (N)) \ , \nonumber
\ea
where $\chi (N)$ denotes the Euler class of the normal bundle.
Letting
\ba
 Y_4^{(+)} &=& - {1 \over 2} \, tr G_1^2 + {d_1 \over 96} \, tr R^2 -
{d_1 \over 48} \, tr N^2 \ , \nonumber \\ 
 Y_4^{(-)} &=& - {1 \over 2} \, tr G_3^2 + {d_3 \over 96} \, tr R^2 -
{d_3 \over 48} \, tr N^2 \ , \label{B063}
\ea
where $G_1$ and $G_3$ describe the ${\rm USp}(d_1) \times {\rm USp}(d_3)$ 
D5-brane gauge fields, 
the residual anomalies on the D1 and D5 branes are compensated by the 
Wess-Zumino terms
\ba
S_{WZ} (D1) &=& T_1 \int_{D1} \left[ d_1 (B_2^{(1)} + B_2^{(2)})
+ d_3 (B_2^{(1)} - B_2^{(2)}) \right] \ , \nonumber \\
S_{WZ} (D5) &=& T_5 \int_{D5} \left[ {d_1 \over 2} \, (B_6^{(1)} + B_6^{(2)})
+ {d_3 \over 2} \, (B_6^{(1)} - B_6^{(2)}) \right] \nonumber \\
&&+\, T_1 \int_{D5} \left[ Y_4^{(+)} (B_2^{(1)} + B_2^{(2)}) +
 Y_4^{(-)} (B_2^{(1)} - B_2^{(2)}) \right] \ . \label{B064}
\ea 
Notice that, since the D5 gauge groups are symplectic, the number of
D5 branes is actually $d_3/2$, as neatly 
reflected in the Wess-Zumino couplings above.
In verifying the anomaly cancellation for the D5 brane, one needs
again the relation between the $\delta$ function on its
world volume and the Euler class of the normal bundle \cite{bott}
\be
\delta (D5)|_{D5} = \chi (N) \ . 
\ee

It is interesting to examine the spectrum of the 
D string, since this could provide hints for the strong
coupling limit of this orientifold. This issue was already considered in
\cite{bg}. As we have seen, there are two types of D1 branes, that differ
in the spectra of the corresponding D1-D9 states. For the first,
there are Majorana-Weyl fermions in the fundamental of
${\rm SO}(n_3)$ and Majorana-Weyl fermions of opposite chirality in the 
fundamental
of ${\rm SO}(n_4)$, while for the second 
$n_{3,4}$ are to be replaced by $n_{1,2}$. For the first type the
central charge, $c=10+n_4/2$, actually becomes critical
if $n_4=32$, and therefore if $n_1=n_2=0$. One is thus tempted, 
following \cite{bg}, to relate the S-dual
of the orientifold with the gauge group ${\rm SO}(32)^2$ to the bosonic string
compactified on an ${\rm SO}(32)$ lattice. An additional argument in favor of this
conjecture is the presence of charged D5 branes in the orientifold, such
that for a single five-brane the 
gauge group is SU(2), since in the S-dual theory this could become   
the NS five-brane of the ten-dimensional bosonic theory with 
gauge group ${\rm SO}(32) \times {\rm SO}(32)$. Compactifications on
group lattices of this type play a central role in the scenario proposed
by Englert {\it et al} to relate all ten-dimensional fermionic strings 
to the bosonic string \cite{englert}.
However, the orientifold contains a second set
of D5 and D1 stable (charged) branes, whose role in the S-dual theory
is not clear.

The orientifold (\ref{B4}) also contains uncharged D-branes. 
The D7, D3 and D$(-1)$ branes are in this case $\Omega$-invariant combinations
of the form $(+,+)+(-,-)$ and $(+,-)+(-,+)$. Since $\Omega$ 
interchanges branes and antibranes of the parent 0B theory, 
one expects unitary gauge groups for
each invariant brane-antibrane configuration, and 
indeed the corresponding open-string amplitudes read
\ba
{\cal A}_{pp} &=& (d_1 {\bar d}_1 + d_2 {\bar d}_2) (V_{p-1}
O_{9-p}+O_{p-1}V_{9-p}) \nonumber \\
&& +\, ({d_1^2 + {\bar d}_1^2 + d_2^2 + {\bar d}_2^2 \over
2}) (O_{p-1} O_{9-p}+ V_{p-1}V_{9-p}) \nonumber \\
&& - \, (d_1 {\bar d}_2+{\bar d}_1 d_2) 
(S_{p-1} S_{9-p}+ C_{p-1} C_{9-p}) \nonumber \\
&& - \, (d_1 d_2+{\bar d}_1 {\bar d}_2) 
(S_{p-1} C_{9-p}+ C_{p-1} S_{9-p}) \ , \nonumber \\
{\cal M}_p &=& -{d_1+{\bar d}_1 + d_2 + {\bar d}_2 \over 2} \, \epsilon\,  
({\hat O}_{p-1}{\hat O}_{9-p}+
{\hat V}_{p-1}{\hat V}_{9-p}) \ ,  \label{B7} 
\ea
where the sign $\epsilon$, determined again by the $P$ matrix, is $+1$ 
for D7 and D$(-1)$ branes and $-1$ for D3 branes. The
resulting gauge groups are therefore ${\rm U}(d_1) \times {\rm U}(d_2)$, and
the open tachyon
can be eliminated for D7 branes with gauge group ${\rm U}(1) \times 
{\rm U}(1)$,
obtained if $d_1=d_2=1$, with a non-chiral fermion spectrum, or for 
a single D7 brane
with a ${\rm U}(1)$ gauge group and no fermions in the spectrum, but not for
the D3 brane, that is thus unstable.
The D7-D9 spectrum can be derived from the amplitude
\ba
{\cal A}_{79} &\!\!=\!\!& (n_1 d_1 \!+\! n_2 {\bar d}_1 \!+\! n_3 d_2 \!+\! 
n_4 {\bar d}_2) (O_{6} S_2\!+\! V_{6}
C_{2}) \nonumber \\
&&+\, (n_1 {\bar d}_1 \!+\! n_2 d_1 \!+\! n_3 {\bar d}_2 \!+\!  
n_4 d_2) (O_{6} C_2\!+\! V_{6} S_{2}) \nonumber \\ 
&&- \, (n_1 \bar{d}_2 + n_2 d_2 + n_3 \bar{d}_1 + n_4 d_1) \ (C_{6} O_2+ S_{6}
V_{2}) \nonumber \\
&&-\,  (n_1 d_2 + n_2 \bar{d}_2 + n_3 d_1+  n_4 \bar{d}_1) \ (S_{6} O_2+ C_{6}
V_{2}) \ , \label{B07} 
\ea
that contains tachyonic modes,
while the D3-D9 amplitude could be obtained from (\ref{B07}) interchanging the
two fermion chiralities $S$ and $C$. For the symplectic orientifolds, 
the roles of D3 and D7 are interchanged, and a
single D3 is now completely free of tachyons, that are also absent
in the D3-D9 sector.
 
The model contains additional uncharged D$p$ branes, with $p=0,2,4,6,8$. The
orientifold projection acts directly on the Chan-Paton factors of the
parent 0B uncharged branes ${\rm D}p_{+}$ and ${\rm D}p_{-}$, and therefore
one can anticipate the presence of orthogonal or symplectic gauge 
groups in all these cases.
If the Chan-Paton multiplicities of  the invariant 
${\rm D}p_{+}$ and ${\rm D}p_{-}$ combinations are denoted by
$d_1$ and $d_2$, the resulting annulus and M\"obius amplitudes read
\ba
{\cal A}_{pp} &\!=\!& {d_1^2 \!+\! d_2^2 \over 2} \ (O_{p-1}\!+\!
V_{p-1})(O_{9-p} \!+\! V_{9-p}) \!-\! 2 d_1 d_2 \ S'_{p-1} S'_{9-p} \ , 
\label{B8} \\ 
{\cal M}_p &\!\!=\!\!& - {d_1\!+\!d_2 \over \sqrt{2}} \left[ \sin{(p\!-\!5) 
\pi \over 4}
({\hat O}_{p\!-\!1}{\hat O}_{9\!-\!p} \!+\! {\hat V}_{p\!-\!1}
{\hat V}_{9\!-\!p}) \!+\! \cos{(p\!-\!5) \pi \over 4}
({\hat O}_{p\!-\!1}{\hat V}_{9\!-\!p}\!-\! {\hat V}_{p\!-\!1}
{\hat O}_{9\!-\!p})  \right] , \nonumber 
\ea
in terms of the non-chiral $S'$ characters defined in (\ref{a01}). 
The D6 and D4 branes have gauge groups ${\rm SO}(d_1) \times {\rm SO}(d_2)$, and the
open tachyons can be eliminated for the D6 brane if $d_1=d_2=1$,
leaving a spectrum without a residual gauge symmetry. 
For D0, D2 and D8 branes, the gauge groups are ${\rm USp}(d_1) \times
{\rm USp}(d_2)$, but these branes are
unstable, since no configuration can eliminate their open-string
tachyons. As usual, for the symplectic orientifolds the orthogonal and
symplectic groups are interchanged for all these lower-dimensional branes.

Finally, the D$p$-D9 spectrum manifests itself in the amplitude
\ba
{\cal A}_{p9} &=& [(n_1+n_2) d_1 + (n_3+n_4) d_2] \ (O_{p-1}+V_{p-1}) \ 
S'_{9-p} 
\nonumber \\
&&- \,  [ (n_3+n_4) d_1 + (n_1+n_2) d_2]  \ S'_{p-1} \ (O_{9-p} + V_{9-p}) \ , 
\label{B08}
\ea
that, as usual, contains tachyons for $p > 5$.
 
The third 0B orientifold of \cite{as95} is determined by the projection 
$\Omega \times (-1)^{G_L}$, so that its Klein
bottle amplitude reads
\ba
{\cal K} &=& {1 \over 2} \ (O_8+V_8+S_8+C_8) \ , \nonumber \\
{\tilde {\cal K}} &=&  2^5 \ O_8 \ . \label{B9}
\ea  

As in the previous case, $\tilde{\cal K}$ does not introduce any R-R tadpoles, 
and therefore the
model is consistent even without introducing D9 branes. 
The orientifold
projection now keeps the two R-R 0-forms and an unconstrained
R-R four-form, and therefore
the model contains two types of 
charged D3, D7 and ${\rm D}(-1)$ branes.  For $p=-1,3,7$, 
the charged branes of the parent 0B are separately invariant, 
and therefore the projection acts diagonally on their Chan-Paton factors, 
generating orthogonal or symplectic gauge groups. On the other hand, for 
$p=5,9$ the
invariant combinations are $(+,+) + (-,-)$ and
 $(+,-) + (-,+)$, and thus one can again anticipate the occurrence of
uncharged branes with unitary gauge groups. Finally, the two type 0B uncharged branes,
${\rm D}p_{+}, {\rm D}p_{-}$, present for even $p$, are left invariant by
the orientifold projection, and therefore in these cases one expects 
orthogonal or symplectic 
gauge groups. The most general uncharged D9 brane 
spectra are based on the gauge group ${\rm U}(n)
\times {\rm U}(m)$, have tachyons in the antisymmetric representation
of the first unitary factor and in the symmetric representation of the
second, and the relevant D9 amplitudes are \cite{as95}
\ba
{\cal A}_{99} &=& (n {\bar n} + m {\bar m}) V_8 + 
{n^2 + {\bar n}^2 + m^2 + {\bar m}^2 \over 2} O_8 \nonumber \\
&&-\, (n {\bar m} +{\bar n} m) S_8 - (n m +{\bar n} {\bar m} ) C_8 \ , \nonumber \\
{\cal M}_{9} &=&  \frac{1}{2} (n+ {\bar n}- m - {\bar m}) \hat{O}_8
 \ . \label{B09}
\ea
Therefore, as in the previous orientifold, there is no
configuration without D9 open-string tachyons. 
The open string amplitudes for the charged D3 and D7 or D$(-1)$ branes read
\ba
{\cal A}_{pp} &\!=\!& 
{d_1^2 \!+\! d_2^2 \!+\! d_3^2 \!+\! d_4^2 \over 2} \, (V_{p-1}
O_{9-p}\!+\!O_{p-1}V_{9-p}) \!+\! (d_1d_2\!+\!d_3d_4) (O_{p-1} O_{9-p}\!+\!
V_{p-1}V_{9-p}) \nonumber \\
&&-\,  (d_1d_3+d_2d_4) (S_{p-1} S_{9-p}+ C_{p-1} C_{9-p})
-  (d_1d_4+d_2d_3) (S_{p-1} C_{9-p}+ C_{p-1} S_{9-p}) \ , \nonumber \\
{\cal M}_p &=& -\, {d_1 + d_2 - d_3 -d_4 \over 2} \ \epsilon \ ({\hat V}_{p-1}
{\hat O}_{9-p}- {\hat O}_{p-1}{\hat V}_{9-p}) \ , \label{B10}
\ea
where $\epsilon$ is $+1$ for D7 and D$(-1)$ branes and $-1$ for D3 branes, so that     
the resulting gauge groups are ${\rm SO}(d_1) \times {\rm SO}(d_2) \times 
{\rm USp}(d_3) \times {\rm USp}(d_4)$ and  ${\rm USp}(d_1) \times 
{\rm USp}(d_2) \times {\rm SO}(d_3)
\times {\rm SO}(d_4)$ in the two cases. As in the parent 0B model, 
there are {\it tachyon-free configurations}, obtained including 
the two types of charged branes but not their
antibranes, {\it i.e} for $d_2=d_4=0$, with gauge 
groups ${\rm SO}(d_1) \times {\rm USp}(d_3)$ for the D7 and 
D$(-1)$ branes and 
${\rm USp}(d_1) \times {\rm SO}(d_3)$ for the D3 branes. Moreover,
configurations with
equal numbers $d$ of $(+,+)$ and $(+,-)$ branes, with gauge groups ${\rm SO}(d)
\times {\rm USp}(d)$, have no brane-brane interactions at the classical 
level.  
The D3 branes are particularly interesting: as for the $0'$B model, 
their spectra become conformal in the large-$d$ limit, and the AdS/CFT 
correspondence  \cite{maldacena} can thus be extended to them even in 
the absence of supersymmetry, proceeding as in \cite{klebanov}. 
The D3-D9 spectrum can be derived from the amplitude
\ba
{\cal A}_{39} &\!=\!& (n d_1\!+\! {\bar n} d_2 \!+\! m d_3 \!+\! {\bar m} d_4) \ 
(O_{2} S_6 \!+\!V_{2} C_{6}) \nonumber \\ &&+\,
(n d_2\!+\! {\bar n} d_1 \!+\! m d_4 \!+\! {\bar m} d_3) \ 
(O_{2} C_6 +V_{2} S_{6}) \nonumber \\
&&-\,  (n d_3+ {\bar n} d_4 + m d_1 + {\bar m} d_2) \ 
(S_{2} O_6 +C_{2} V_{6}) \nonumber \\
&&-\, (n d_4+ {\bar n} d_3 + m d_2 + {\bar m} d_1) \ 
(S_{2} V_6 +C_{2} O_{6}) \ , \label{B010}
\ea
and contains massless fermions in bi-fundamental representations,
while the D7-D9 amplitude is very similar, aside from 
the interchange of the two spinor chiralities.

As in the previous orientifold, the lower-dimensional charged (D3 and D7) 
branes have an anomaly inflow from the D9 branes. In this case,
the anomaly polynomial is given by
\be
I_{12} = X_2^{(+)} X_{10}^{(-)} +  X_2^{(-)} X_{10}^{(+)}+ 
X_6^{(+)} X_6^{(-)} \ , \label{B0100} 
\ee
with
\ba
X_2^{(+)} &=&- i \ tr F_2 \ , \qquad 
X_2^{(-)} =  2 i \ tr F_1 \ , \nonumber \\
X_6^{(+)} &=& - {i \over 3} \, tr F_2^3 \ , \qquad 
X_6^{(-)} =  {i \over 6}  \, tr F_1^3 \ , \nonumber \\
X_{10}^{(+)} &=& -  {i \over 120} \, tr F_2^5 + {i \over 288} \, tr R^2 
\, tr F_2^3
- i \, {7p_1^2 - 4 p_2 \over 2 \times 5760} \, tr F_2 \ , \nonumber \\
X_{10}^{(-)} &=&  {i \over 60} \, tr F_1^5 -{i \over 144} \, tr R^2 \, tr F_1^3
+i \, {7p_1^2 - 4 p_2 \over 5760} \, tr F_1 \ ,
\label{B101} 
\ea
where $F_1$, $F_2$ describe the D9  ${\rm U}(m) \times {\rm U}(n)$ field
strengths. 
The anomaly polynomials for the D3 and D7 branes are
\ba
I_6 &=& d_3 (-X_6^{(-)}+{{d_1}\over{4}}\chi(N)) + {d_1 \over
2}(X_6^{(+)}+{d_3 \over 2} \chi(N))
 + X_2^{(+)} Y_4^{(+)} +
X_2^{(-)} Y_4^{(-)} \ , \nonumber \\ 
I_{10} &=& {d_3 \over 2}(- X_{10}^{(-)}+{{N}\over{2}}Z_8^{(+)}) + 
d_1 (X_{10}^{(+)}
-{{N}\over{2}}Z_8^{(-)}) + (X_6^{(+)}-{{N}\over{2}}Z_4^{(-)})Z_4^{(+)}\nonumber\\
&&+\,
 (X_6^{(-)}-{{N}\over{2}}Z_4^{(+)}) Z_4^{(-)} +
 ( X_2^{(+)}+{{d_3}\over{4}}N) Z_8^{(+)} + (X_2^{(-)}- {{d_1}\over{2}}N)
  Z_8^{(-)} \ , \label{B102} 
\ea
where 
\ba
Y_4^{(+)} &=& { 1 \over 2} \, tr G_1^2 - {d_1 \over 48} \, (tr R^2-tr N^2) \ , 
\nonumber \\ 
Y_4^{(-)} &=& -{ 1 \over 4} \, tr G_3^2 + {d_3 \over 96} \, 
(tr R^2- tr N^2) \ ,
\label{B103}  \\
Z_4^{(+)} &=& {1 \over 4} \, tr G_1^2 - {d_1 \over 48} N^2 \ , \qquad 
Z_4^{(-)} = -{1 \over 2} \, tr G_3^2  + {d_3 \over 24} N^2 \ , \nonumber \\ 
Z_8^{(+)} &=& {1 \over 24} \, tr G_1^4 -{1 \over 48} tr G_1^2 N^2 + {d_1
\over 1920} N^4 - {1 \over 96} \, tr R^2 \, (tr G_1^2-{d_1 \over 12} N^2)  +
d_1 \, {7p_1^2-4 p_2 \over 2 \times 5760} \ , \nonumber \\
Z_8^{(-)} &=& -{1 \over 48} \, tr G_3^4 +{1 \over 96} tr G_3^2 N^2
- {d_3 \over 3840} N^4+ {1 \over 192} \, tr R^2 \, (tr G_3^2-{d_3 \over
12} N^2) - d_3 \, {7p_1^2-4 p_2 \over 4 \times 5760}\ . \nonumber 
\ea 
In (\ref{B102}), $\chi(N)$ denotes the Euler class of the normal
bundle for the D3 brane, and $N$ denotes the curvature of the U(1) normal
bundle for the D7 brane.
  
The resulting Wess-Zumino couplings are then
\ba
S_{WZ} (D3) &=& T_3 \int_{D3} \left[ {d_1 \over 2} \, (B_4^{(1)} + B_4^{(2)})
+ d_3 (B_4^{(1)} - B_4^{(2)}) \right] \nonumber \\
&+& T_{-1} \int_{D3} \left[ (B_0^{(1)} + B_0^{(2)}) Y_4^{(+)} +
 (B_0^{(1)} - B_0^{(2)}) Y_4^{(-)} \right] \ , \nonumber \\
S_{WZ} (D7) &=& T_7 \int_{D7} \left[ d_1 (B_8^{(1)} + B_8^{(2)})
+ {d_3 \over 2} \, (B_8^{(1)} - B_8^{(2)}) \right] \nonumber \\
&+& T_{3} \int_{D7} \left[ (B_4^{(1)} + B_4^{(2)}) Z_4^{(+)} +
 (B_4^{(1)} - B_4^{(2)}) Z_4^{(-)} \right] \nonumber \\
&+& T_{-1} \int_{D7} \left[ (B_0^{(1)} + B_0^{(2)}) Z_8^{(+)} +
(B_0^{(1)} - B_0 ^{(2)}) Z_8^{(-)} \right] \ . \label{B104}
\ea 
Notice that, again, the number of symplectic-like
branes is actually $d_1/2$ in the D3 case and $d_3/2$ in the D7
case.
In (\ref{B104}), $B_0^{(1,2)}$ are the two R-R
zero-forms, $B_8^{(1,2)}$ are their duals, $B_4^{(1)}$ is the self-dual
four-form and $B_4^{(2)}$ the anti self-dual four-form. Once more,
in verifying the anomaly cancellation induced by the Wess-Zumino terms, 
it is crucial to use \cite{bott} 
\be 
\delta(D3)|_{D3}= \chi(N) \ ,\quad  \delta(D7)|_{D7}=N \ . \label{B106}
\ee

There are also uncharged D5 and D1 branes,
with open string amplitudes
\ba
{\cal A}_{pp} &=& (d {\bar d} + q {\bar q}) (V_{p-1}
O_{9-p}+O_{p-1}V_{9-p}) + {d^2 + {\bar d}^2 + q^2 + {\bar q}^2 \over
2}\, (O_{p-1} O_{9-p}+ V_{p-1}V_{9-p}) \nonumber \\
&&-\, (d {\bar q}+{\bar d} q) 
(S_{p-1} S_{9-p}+ C_{p-1} C_{9-p}) - (d q+{\bar d} {\bar q}) 
(S_{p-1} C_{9-p}+ C_{p-1} S_{9-p}) \ , \nonumber \\
{\cal M}_p &=& - \, \epsilon \, {d+{\bar d} - q - {\bar q} \over 2} \, ({\hat O}_{p-1}{\hat O}_{9-p}+
{\hat V}_{p-1}{\hat V}_{9-p}) \ ,  \label{B11} 
\ea
where $\epsilon$ is $+1$ for the D5 branes and $-1$ for the D1 branes, 
with gauge groups ${\rm U}(d) \times {\rm U}(q)$, 
as previously
anticipated. The open-string tachyon
can be eliminated for a single D1 brane or for a single D5 brane,
with a U(1) gauge group and no massless
space-time fermions. The D5-D9 amplitude is
\ba
{\cal A}_{59} &\!=\!& (n {\bar d}\!+\! {\bar n} d \!+\! m {\bar q} \!+\! {\bar m} q) \ 
(O_{4} S_4 \!+\!V_{4} C_{4}) \!+\!
 (n d\!+\! {\bar n} {\bar d} \!+\! m q \!+\! {\bar m} {\bar q}) \ 
(O_{4} C_4 \!+\!V_{4} S_{4}) \label{B011} \\
&&\!\!\!\!-\! (n q\!+\! {\bar n} {\bar q} \!+\! m d \!+\! {\bar m} {\bar d}) \ 
(S_{4} O_4 \!+\!C_{4} V_{4}) \!-\!
 (n {\bar q}\!+\! {\bar n} q \!+\! m {\bar d} \!+\! {\bar m} d) \ 
(S_{4} V_4 \!+\!C_{4} O_{4}) \ , \nonumber
\ea
while the D1-D9 amplitude could be obtained from it
interchanging the two spinor chiralities.
   
Finally, there are uncharged D$p$ branes with $p=0,2,4,6,8$, whose
open string amplitudes are
\ba
{\cal A}_{pp} &\!=\!& {d_1^2 + d_2^2 \over 2} \ (O_{p-1}+ V_{p-1})(O_{9-p}+
V_{9-p}) - 2 d_1 d_2 \ S'_{p-1} S'_{9-p} \ , \label{B12} \\ 
{\cal M}_p &\!\!=\!\!& \! {d_2\!-\! d_1 \over \sqrt{2}} \left[ 
\cos{(p\!-\!5) \pi \over 4}
({\hat O}_{p-1}{\hat O}_{9-p} \!+\! {\hat V}_{p-1}{\hat V}_{9-p}) \!+\!
 \sin{(p\!-\!5) \pi \over 4}
({\hat V}_{p-1}{\hat O}_{9-p}\!-\! {\hat O}_{p-1}{\hat V}_{9-p})  \right] 
 , \nonumber
\ea
so that the resulting gauge groups for the $D_{+}$ and  $D_{-}$ branes
are ${\rm SO}(d_1)$ and ${\rm USp}(d_2)$. The
open-string tachyons can be eliminated for the D2 and D6 branes,
leaving no residual gauge symmetry. On the other hand, the remaining
D0, D4 and D8 branes are all unstable.
\section{The D-branes of the 0A orientifold}

One can also derive the charged and
uncharged brane spectrum of the 0A orientifold, whose D9
structure was first displayed in \cite{bianchias}. As 
we shall see, this model contains non-tachyonic charged branes of lower
dimensionalities with full-fledged non-Abelian gauge groups.

The 0A orientifold is obtained supplementing the torus amplitude
of \cite{dhsw}
\be
{\cal T} = |O_8|^2 +|V_8|^2 + S_8 {\bar C}_8 + C_8 {\bar S}_8 
\ , \label{0A3}
\ee
with the Klein-bottle amplitudes \cite{bianchias}
\ba
{\tilde {\cal K}} &=& {2^5 \over 2} (O_8+V_8) \ , \nonumber \\
{\cal K} &=& {1 \over 2} (O_8+V_8) \ , \label{0A4}
\ea 
related as usual by an $S$ transformation.
In this case the orientifold projection $\Omega$ can not eliminate the 
closed-string tachyon, that will consequently couple to the open sector.
It interchanges the  two types
of R-R $n$-forms ($n=1,3,5,7,9$) in the parent $0A$ model, here denoted by
$A_n$ and $A'_n$, according to
\be
\Omega A_n=(-1)^{(n+1)(n+2)/2}A'_n \ ,  \label{OM}
\ee 
as can be deduced from the corresponding $\gamma$ matrices.
This novel feature compared to the 0B or IIB cases, where $\Omega$ acts 
diagonally on the R-R fields, reflects the nature
of the 0A model, that is described by a {\it non-diagonal} modular
invariant. Notice that the orientifold projection of this model
is not connected by T-duality to any of the three 0B orientifold
projections discussed in the previous section. 

As one can see from $\tilde{\cal K}$, there is no induced 
R-R tadpole, and therefore also in this case one could refrain from 
including a D9 open sector. One has, however, the option
of including it, as in \cite{bianchias}, and this contains precisely the 
two types of uncharged branes already discussed in Section 3, albeit with
suitably projected Chan-Paton assignments. The D9-D9 amplitude is thus
described by
\ba
{\tilde {\cal A}}_{99} &=& {2^{-5} \over 2} \left[ (n_1+n_2)^2 
V_8 +  (n_1-n_2)^2 O_8 \right] \ , \nonumber \\
{\cal A}_{99} &=& {n_1^2 + n_2^2 \over 2} \ (O_{8}+ V_{8}) - \ 
n_1 n_2 (S_{8}+ C_{8}) 
 \ , \label{0A5}  
\ea
where the branes of the first type, with Chan-Paton multiplicity
$n_1$, have a positive coupling to the closed-string tachyon, while those
of the second type, with Chan-Paton 
multiplicity $n_2$, have a negative coupling to it.
The corresponding D9-O9 M\"obius amplitude is then
\ba
{\tilde{\cal{M}}}_9 &=& -\, (n_1-n_2) {\hat O}_8 - 
(n_1+n_2) {\hat V}_8 \ , \nonumber \\
{\cal M}_9 &=& -\, {n_1+n_2 \over 2} \, {\hat V}_8 + 
{n_1 - n_2 \over 2} \, {\hat O}_8 
 \ , \label{0A6}
\ea
and the D9 gauge group is therefore ${\rm SO}(n_1) \times {\rm SO}(n_2)$, with
tachyons in the $(n_1(n_1+1)/2,1)+ (1,n_2(n_2-1)/2)$ representations
and Majorana fermions in the bi-fundamental representation. 
Since the dilaton tadpole condition $n_1+n_2=32$ 
only eliminates the need for
a background redefinition \cite{fs} of the type discussed explicitly 
in \cite{dm2}, also in this case one has the additional 
option of reversing both $n_1$ and $n_2$, thus 
replacing the ${\rm O}9_+$ planes with
${\rm O}9_-$ ones, with opposite tachyon and dilaton couplings.
In both cases the
D9 branes are unstable and can decay into the vacuum.

This model has also charged D$p$ branes, with $p$=0,2,4,6,8, that 
can be deduced from the brane configurations of the parent $0A$
that 
are invariant under $\Omega$. Here the signs in eq. (\ref{OM}) play a
crucial role, and indeed when $\Omega A=A'$, the configurations
invariant under $\Omega$
are of the type $d_1  {\rm D}p_1 + d_2 \overline{\rm D}p_1$ 
or $m({\rm D}p_2+{\overline{\rm D}p}_2)$, 
while when $\Omega A=-A'$ they are of the type
$d_1 {\rm D}p_2 + d_2 {\overline{\rm D}}p_2$ and $m({\rm D}p_1+{\overline{\rm D}p}_1)$. 
For $p=0,2,4,6,8$, the D$p$-D$p$ annulus amplitudes are thus
\ba
\tilde {\cal A}_{pp}&=&{{2^{-(p+1)/2}}\over{4}}\bigl[
(d_1+d_2+m+\bar m)^2 (V_{p-1} O_{9-p} +  O_{p-1}
V_{9-p})  \\
&&+\, (d_1+d_2-m-\bar m)^2 (O_{p-1}O_{9-p}+V_{p-1} V_{9-p}) \nonumber \\
&&- \, |d_1-d_2-m+\bar m|^2  \, 
(S_{p-1} + C_{p-1})(S_{9-p} + C_{9-p}) \bigr] 
\ , \nonumber \\
{\cal A}_{pp} &=& ({d_1^2+d_2^2 \over 2}+m {\bar m})
 (O_{p-1}V_{9-p}+ V_{p-1}O_{9-p}) \nonumber \\
&&+\, (d_1 d_2+{{m^2+\bar m^2}\over{2}})
 \ (O_{p-1}O_{9-p}+ V_{p-1} V_{9-p}) 
 -\, (d_1+d_2)(m+\bar m) S'_{p-1} S'_{9-p} \ , \nonumber
\ea
where we have expressed the open spinor content, $1/2(S_8+C_8)$, in terms
of the single odd-dimensional $S'$ spinors. For $p=0,4$ and $8$ the 
M\"obius amplitudes are
 \ba
 {\tilde {\cal M}}_p &=&
- \, {{d_1+d_2 +m+\bar m} \over {\sqrt{2}}} \   
({\hat V}_{p-1}{\hat O}_{9-p}- {\hat O}_{p-1}{\hat V}_{9-p}) \label{0A10} \\
&&-\, {{d_1+d_2 -m-\bar m} \over {\sqrt{2}}}
({\hat O}_{p-1}{\hat O}_{9-p}+ {\hat V}_{p-1}{\hat V}_{9-p}) \ ,
\nonumber \\
 {\cal M}_p &=&
 {d_1+d_2 \over 2} \ \epsilon \ 
({\hat O}_{p-1}{\hat V}_{9-p}- {\hat V}_{p-1}{\hat O}_{9-p})
-\, {{m+\bar m}\over{2}} \ \epsilon \ 
({\hat O}_{p-1}{\hat O}_{9-p}+ {\hat V}_{p-1}{\hat V}_{9-p}) \
, \nonumber
\ea
where $\epsilon$ is $+1$ for $p=0,8$ and $-1$ for $p=4$.
The resulting gauge groups are therefore ${\rm SO}(d_1) \times
{\rm SO}(d_2) \times {\rm U}(m)$ for $p=0,8$ and ${\rm USp}(d_1) \times {\rm USp}(d_2)
\times {\rm U}(m)$ for $p=4$ in the
orthogonal 0A orientifolds, and {\it vice versa} in the symplectic ones.
Notice that for $m=0$ and $d_1=0$ (or $d_2=0$) these spectra contain
{\it no open-string tachyons}. 
In a similar fashion, for $p=2,6$ the M\"obius amplitudes are
 \ba
 {\tilde {\cal M}}_p &=&
- \, {{d_1+d_2 +m+\bar m} \over {\sqrt{2}}} \   
({\hat V}_{p-1}{\hat O}_{9-p}- {\hat O}_{p-1}{\hat V}_{9-p}) \nonumber \\
&&+\, {{d_1+d_2 -m-\bar m} \over {\sqrt{2}}}
({\hat O}_{p-1}{\hat O}_{9-p}+ {\hat V}_{p-1}{\hat V}_{9-p}) \ ,
\nonumber \\
 {\cal M}_p &=&
 {d_1+d_2 \over 2} \ \epsilon \ 
({\hat O}_{p-1}{\hat V}_{9-p}- {\hat V}_{p-1}{\hat O}_{9-p})
\nonumber \\
&& -\,{{m+\bar m}\over{2}} \ \epsilon \ 
({\hat O}_{p-1}{\hat O}_{9-p}+ {\hat V}_{p-1}{\hat V}_{9-p}) \
, \label{0A11}
\ea
where $\epsilon$ is $+ 1$ for $p=2$ and $-1$ for
$p=6$. Notice that the closed-string 
tachyon appears in the transverse M\"obius amplitudes (\ref{0A10}) and 
(\ref{0A11})
with opposite signs, consistently with the fact that the branes involved in
the two cases, of types ${\rm D}p_1$ and ${\rm D}p_2$, have opposite
tachyonic couplings. The gauge groups for $p=2$ ($p=6$) are analogous
to those for $p=0,8$ ($p$=4). The $0A$ orientifold has thus a rather unusual 
feature: {\it for the same (even) $p$}, there are both charged branes, 
with orthogonal or symplectic gauge groups, and uncharged ones, 
with unitary gauge groups.
As usual, there are additional states in the D$p$-D9 sector, if 
D9 branes are present, whose annulus amplitudes are
\ba
{\tilde {\cal A}}_{p9} &\!=\!& {2^{-5} \over \sqrt{2} } \bigl\{ 
(n_1 \!-\! n_2) (d_1+d_2-m -\bar{m})
(O_{p-1}O_{9-p} \!-\! V_{p-1}V_{9-p})  \nonumber \\
 && + \, (n_1\!+\!n_2) (d_1+d_2+m +\bar{m}) (V_{p-1}
O_{9-p} \!-\! O_{p-1} V_{9-p}) \bigr\} \ , \nonumber  \\
{\cal A}_{p9} &\!=\!& \bigl[n_1(d_1+d_2) + n_1 \bar{m} + n_2 m\bigr]  
\ (O_{p-1} + V_{p-1}) \ S'_{9-p} \nonumber \\
&& - \, \bigl[n_2(d_1+d_2) + n_1 m + n_2 \bar{m} \bigr] 
\ S'_{p-1} (O_{9-p} + V_{9-p}) \ . \label{0A12}
\ea
There are therefore massless fermions in the
bi-fundamental representation of ${\rm SO}(n_2)$ and of the 
D$p$ gauge group and
scalars (tachyonic for $p > 5$) in the bi-fundamental of
${\rm SO}(n_1)$ and the D$p$ gauge group.

This model also contains uncharged D$p$ branes for
$p=1,3,5,7$, that can be obtained by an orientifold projection from
the ${\rm D}p_{\pm}$ 0A branes discussed in Section 3, and are 
therefore of two types,
depending on the sign of their coupling to the closed-string tachyon.
The D$p$-D$p$ amplitudes 
\ba
{\tilde {\cal A}}_{pp} &\!\!=\!\!& {2^{-(p+1)/2} \over 2} \left[ 
(d_1\!+\!d_2)^2 (V_{p-1} O_{9-p} \!+\!  O_{p-1}
V_{9-p}) \!+\!  (d_1-d_2)^2 (O_{p-1}O_{9-p}\!+\!V_{p-1} V_{9-p}) \right] \ , \nonumber \\
{\cal A}_{pp} &=& {d_1^2 + d_2^2 \over 2} \ (O_{p-1}+ V_{p-1})(O_{9-p}+
V_{9-p})   \nonumber \\ 
&& - \, d_1 d_2 \, (S_{p-1}+ C_{p-1}) (S_{9-p}+ C_{9-p})  
 \label{0A13}  
\ea
are obtained introducing real
Chan-Paton charges $d_1,d_2$ and dimensionally reducing on the D$p$
world-volume the D9-D9 amplitudes (\ref{0A5}), while the M\"obius 
amplitudes are
\ba
{\tilde {{\cal M}_p}} &\!=\!& -  
(d_1\!-\!d_2) \ ({\hat O}_{p-1}{\hat O}_{9-p}\!+\! 
{\hat V}_{p-1}{\hat V}_{9-p}) \!-\! (d_1\!+\! d_2) \ ({\hat
V}_{p-1}{\hat O}_{9-p} \!-\! {\hat O}_{p-1}{\hat V}_{9-p}) \nonumber \\
{\cal M}_p &\!=\!&  {d_1 - \epsilon' d_2 \over 2} \, \epsilon\,  
({\hat O}_{p-1}{\hat O}_{9-p} \!+\! {\hat V}_{p-1}{\hat V}_{9-p}) \!+\!
 {d_1 \!+\! \epsilon' d_2 \over 2} \, \epsilon \,
({\hat O}_{p-1}{\hat V}_{9-p}\!-\! {\hat V}_{p-1}{\hat O}_{9-p}) \
 \, . \label{0A14}
\ea    
The signs $(\epsilon, \epsilon')$ are $(+1,+1)$ for the D1
branes,  $(+1,-1)$ for the D3 branes, $(-1,+1)$ for the D5 branes\
and $(-1,-1)$ for the D7 and D$(-1)$ branes. In the orthogonal orientifolds, 
the resulting gauge groups are 
therefore ${\rm SO}(d_1) \times {\rm SO}(d_2)$ for the D1
branes, ${\rm SO}(d_1) \times {\rm USp}(d_2)$ for the D3 branes, ${\rm USp}(d_1) \times
{\rm USp}(d_2)$ for the D5 branes and ${\rm USp}(d_1) \times {\rm SO}(d_2)$ 
for the D7 and D$(-1)$ branes. In the symplectic orientifolds, the
overall sign of ${\cal M}_p$ would be reverted, and therefore the roles of
the D3 and D7 (or D$(-1)$) and D1 and D5  branes would be interchanged.

The tachyons can be eliminated for one D1 brane or for one D7 brane, 
{\it i.e.} if $d_1=0$, $d_2=1$. 
Finally, the D$p$-D9 spectrum is described by 
\ba
{\tilde {\cal A}}_{p9} &\!=\!& 2^{-5} \ \Big[(n_1-n_2) (d_1-d_2) 
(O_{p-1}O_{9-p} - V_{p-1}V_{9-p}) \nonumber \\
&&+\, (n_1+n_2)(d_1+d_2) (V_{p-1}
O_{9-p}\!-\! O_{p-1} V_{9-p}) \Big] \  , \nonumber \\
{\cal A}_{p9} &=& (n_1d_1+n_2d_2) \ (O_{p-1} + V_{p-1}) (S_{9-p}+ C_{9-p})
\nonumber \\ &&-\,
(n_1d_2+n_2d_1) (S_{p-1}+C_{p-1}) (O_{9-p}+ V_{9-p}) 
 \ , \label{0A15}  
\ea 
where, as usual, new tachyons appear if
$p > 5$. All the D$p$-brane spectra in the 0A orientifold are non-chiral.

As already mentioned, a single D1 brane ($d_1=0$, $d_2=1$) is stable and
contains two-dimensional Majorana 
fermions in the fundamental representation of the ${\rm SO}(n_1)$ D9
gauge group. The resulting central charge $c_{L,R} = 10 + n_1/2$ reaches its
critical value 26 for $n_1=32$, when the D9 gauge group becomes
${\rm SO}(32)$. It is interesting to notice that this value is 
singled out in ten dimensions if the tachyon coupling is also
eliminated in $\tilde{\cal A}$. All this suggests
that the strong coupling limit of the 0A orientifold with
${\rm SO}(32)$ gauge group can be related to the quantization of this 
critical ten-dimensional
closed string, with a gauge group
${\rm SO}(32) \times {\rm SO}(32)$. On the other hand, the 0A 
orientifold suggests
that the left and right-moving world-sheet fermions should generate 
the same gauge
group, and it is then natural to conjecture that the S-dual of
this particular
0A orientifold is actually an orientifold of the above-mentioned
critical closed string theory, where the two gauge factors are
identified, so that one finally obtains 
an ${\rm SO}(32)$ gauge group. Its modular invariant torus amplitude 
is
\be
{\cal T} = |O_{32}|^2 + |V_{32}|^2 + |S_{32}|^2 + |C_{32}|^2 \ ,
\ee
to be combined with the Klein-bottle projection
\be 
{\cal K} = {1 \over 2} (O_{32} + V_{32} - S_{32}-  C_{32}) \ .
\label{0A16}
\ee  
The resulting massless spectrum comprises the
graviton, the dilaton and the ${\rm SO}(32)$ gauge bosons, together
with scalars in a 4-index
representation of the gauge group.  The low-lying spectra of the two
theories do not exactly coincide, since the 0A orientifold has a
one-form and a three-form not present in the
orientifold (\ref{0A16}) with no open sector. In addition, the charged 
massless scalars present in (\ref{0A16}) have no counterpart in the 0A orientifold.
However, (\ref{0A16}) contains a singlet
tachyon of mass $m^2 = -4 / \alpha'$ and additional tachyons in the symmetric
representation of ${\rm SO}(32)$, of mass $m^2 = -1 / \alpha'$ that,
interestingly enough, match the spectrum of
corresponding closed and open tachyons of the 0A orientifold.  

\section{Comments on fractional branes}

Orbifold projections can also lead to the appearance, 
at orbifold fixed points, of fractional branes, that
carry new types of (twisted) R-R charges, lack the
moduli associated to their displacements, 
and have tensions smaller than those of the
branes that can move to the bulk. In addition, the O-planes themselves
can acquire twisted R-R charges as a result of $Z_N$ projections with
$N \neq 2$.
Fractional branes present themselves in a number of orbifold instances, 
and in particular in orientifolds, where twisted R-R forms play
a central role in the
generalized Green-Schwarz mechanism \cite{ggs}. For instance, overall neutral 
combinations of branes with twisted R-R couplings of this type 
are present in the supersymmetric models of \cite{bianchias}, that are
$Z_2$ orbifold reductions with a quantized $B_{ab}$
\cite{descendants}, as in \cite{kakt}, albeit with rational geometric
moduli, and in the ``brane supersymmetry breaking'' $T^4/Z_2$ model of
\cite{bsb}. Further, non-neutral combinations of fractional
branes are present 
in any supersymmetric $Z_N$ orbifold with $N \neq 2$, and for instance 
in \cite{z3}, where they absorb the twisted R-R charge that the
O-planes acquire in these cases.

It is instructive to display an explicit example of such
branes, in the spirit of the preceding sections. To this end, let us
consider the 0'B D3-branes \cite{dm1}
of (\ref{B03}) and (\ref{B033}) at an orbifold singularity in
$R^{1,5}\times R^4 / Z_2$ \footnote{Fractional IIB branes on this
orbifold were previously considered in \cite{dkt}.}. 
The orbifold action breaks the D3 gauge group to 
${\rm U}(d_1)\times {\rm U}(d_2)$, while new twisted tadpoles, inadmissible
for the space-filling background D9 branes, demand that their U(32) gauge 
group break to ${\rm U}(16) \times {\rm U}(16)$. The direct-channel
amplitudes
\ba
{\cal A}_{33} &=& {1 \over 2} \Bigl\{ |d_1+d_2|^2 (V_4 O_4+ O_4
V_4)
-{(d_1+d_2)^2+({\bar d}_1+{\bar d}_2)^2 \over 2} (C_4 S_4 + S_4 C_4) 
\label{c01} \\
&& +|d_1-d_2|^2 (V_4 O_4-O_4V_4)
+ {(d_1-d_2)^2+({\bar d}_1-{\bar d}_2)^2 \over 2} (C_4 S_4 - S_4 C_4)
\Bigr\} \ , \nonumber \\
{\cal M}_3 &=& {1 \over 4} \Bigl\{ (d_1+d_2- {\bar d}_1 - {\bar d}_2 ) 
( {\hat S}_2 {\hat C}_2 {\hat S}_4 + {\hat S}_2 {\hat S}_2 {\hat C}_4- 
{\hat C}_2 {\hat S}_2 {\hat S}_4- {\hat C}_2 {\hat C}_2 {\hat C}_4 )
\nonumber \\
&&-  (d_1+d_2- {\bar d}_1 - {\bar d}_2 ) 
( {\hat S}_2 {\hat C}_2 {\hat S}_4 - {\hat S}_2 {\hat S}_2 {\hat C}_4- 
{\hat C}_2 {\hat S}_2 {\hat S}_4+ {\hat C}_2 {\hat C}_2 {\hat C}_4 )
\Bigr\} \ , \nonumber \\
{\cal A}_{93} &=& ({\bar n}_1 d_1+{\bar n}_2 d_2 ) (O_2C_2C_4+V_2S_2C_4)
+  (n_1 {\bar d}_1+n_2 {\bar d}_2) (O_2S_2C_4+V_2C_2C_4) \nonumber \\
&&+ ({\bar n}_1 d_2+{\bar n}_2 d_1 ) (O_2S_2S_4+V_2C_2C_4)
+  (n_1 {\bar d}_2+n_2 {\bar d}_1) (O_2C_2S_4+V_2S_2S_4) \nonumber \\
&&- (n_1 d_1+n_2 d_2 ) (S_2O_2O_4+C_2V_2O_4)
- ({\bar n}_1 {\bar d}_1+ {\bar n}_2 {\bar d}_2 ) (C_2O_2O_4+S_2V_2O_4)
\nonumber \\
&&-  (n_1 d_2+n_2 d_1) (S_2V_2V_4+C_2O_2V_4)
- ({\bar n}_1 {\bar d}_2+ {\bar n}_2 {\bar d}_1) (C_2V_2V_4+S_2O_2V_4) 
 \nonumber
\ea 
can be obtained enforcing the $Z_2$ orbifold projection in 
(\ref{B03}) and (\ref{B033}), while the D3-D3 closed-channel amplitude
\ba
{\tilde{\cal A}}_{33} &=& {2^{-2} \over 8} \Bigl[ (d_1+d_2+{\bar d}_1
+{\bar d}_2)^2 (V_4O_4+O_4V_4-S_4C_4-C_4S_4) \nonumber \\
&&- \ (d_1+d_2-{\bar d}_1 -{\bar d}_2)^2 (O_4O_4+V_4V_4-S_4S_4-C_4C_4)
\nonumber \\
&&+ \ (d_1-d_2+{\bar d}_1-{\bar d}_2)^2 (O_4C_4+V_4S_4-S_4V_4-C_4O_4)
\nonumber \\
&&- \ (d_1-d_2-{\bar d}_1 +{\bar d}_2)^2 (O_4S_4+V_4C_4-S_4O_4-C_4V_4)
\Bigr] \
 \label{c02}
\ea
encodes the D3
brane couplings to the closed sector, and it is transparent that this
brane configuration indeed
contains $N$ regular branes and $M$ fractional branes with half-tension,
where $N$ and $M$ are defined by \footnote{Actually, $d_1(d_2$) simply 
count the numbers of (anti)branes carrying twisted R-R charges. 
$|d_1-d_2|$ of them can not move off the fixed points, and define the
fractional branes.}
\be
d_1 = N+M \quad , \quad d_2 = N \ . \label{c2}
\ee
In addition, the last two lines in (\ref{c02})
display the couplings of fractional branes to twisted closed
fields, that are clearly reminiscent of similar patterns found in
orientifold vacua.

The resulting massless modes are described by
\ba
{\cal A}_{33,0} + {\cal M}_{33,0} + {\cal A}_{93,0} &=&
(d_1 \bar{d}_1 + d_2 \bar{d}_2) (V_2 O_2 O_4 + O_2 V_2 O_4) + 
(d_1 \bar{d}_2 + d_2 \bar{d}_1) O_2 O_2 V_4 \nonumber \\
&& - (d_1 d_2 + \bar{d}_1 \bar{d}_2 ) (S_2 C_2 S_4 + C_2 S_2 S_4)
\nonumber \\
&&- \frac{d_1(d_1-1) + d_2(d_2-1) + \bar{d}_1(\bar{d}_1+1) + 
\bar{d}_2(\bar{d}_2+1) }{2} S_2 S_2 C_4 \nonumber \\
&& - 
\frac{d_1(d_1+1) + d_2(d_2+1) + \bar{d}_1(\bar{d}_1-1) + 
\bar{d}_2(\bar{d}_2-1) }{2} C_2 C_2 C_4 \nonumber \\
&& + ( \bar{n}_1 d_1 + \bar{n}_2 d_2) O_2 C_2 C_4 +
(n_1 \bar{d}_1 + n_2 \bar{d}_2) O_2 S_2 C_4  \nonumber \\
&& +(\bar{n}_1 d_2 + \bar{n}_2 d_1) O_2 S_2 S_4
+ (n_1 \bar{d}_2 + n_2 \bar{d}_1) O_2 C_2 S_4 \nonumber \\
&& - (n_1 d_1 + n_2 d_2) S_2 O_2 O_4 - (\bar{n}_1 \bar{d}_1 + 
\bar{n}_2 \bar{d}_2) C_2 O_2 O_4 \ , \label{c1}
\ea
and the resulting spectrum is chiral but free of irreducible 
gauge anomalies for any $d_1$ and $d_2$. 
In addition to the $U(d_1) \times U(d_2)$ gauge vectors, the D3-D3 sector
contains two real scalars in the adjoint of the D3 gauge
group, four complex scalars in the $(d_1,\bar{d}_2)$, 
two Dirac fermions in the $(d_1,d_2)$, two Weyl 
fermions of positive chirality in antisymmetric representations 
and two Weyl fermions of negative chirality in
symmetric representations. Moreover,
the D9-D3 spectrum contains pairs of complex scalars in the $(n_1,\bar{d}_2)$, 
$(n_2,\bar{d}_1)$, $(n_1,\bar{d}_1)$ and  $(n_2,\bar{d}_2)$ and Weyl
fermions of positive chirality in the  $(n_1,d_1)$ and $(n_2,d_2)$.
The third 0B orientifold model of (\ref{B9}) allows a similar
construction containing $m$ fractional branes, with a resulting
gauge group SO($n+m$)$\times$SO($n$),
and/or $s$ fractional branes, with a resulting gauge group
USp($r+s$)$\times$USp($r$). In particular, the force-free configurations, 
with $2 n + m = 2 r + s$, are suitable for applications to the 
gauge-gravity correspondence.

Fractional branes are actually a generic feature of orbifolds of
conformal theories, a
phenomenon neatly illustrated by the SU(2) WZW models. For 
the diagonal $A$-series, a nice geometrical interpretation 
has recently emerged,
in terms of D2 branes on special SU(2) conjugacy classes \cite{schom}, 
and the level-$k$ model has $(k+1)$ such branes, corresponding to
its $(k+1)$ Cardy states, whose couplings to states of half-odd-integer
isospin mimic the usual R-R charges. There are, however, additional
non-diagonal modular invariants, with more peculiar properties.
This is the case, in particular, for the whole $D_{odd}$ series of \cite{ciz},
that presents an amusing extension of the boundary symmetry
first noticed in \cite{completeness}. The simplest example of this type, 
the $D_5$ model, occurs at level $k=6$ and can be obtained from 
the $A_6$ model
\be
{\cal T}_{A_6} = |\chi_1|^2 + |\chi_3|^2 + |\chi_5|^2 + |\chi_7|^2 +
|\chi_4|^2 + |\chi_2|^2 + |\chi_6|^2 \ ,
\label{c4}
\ee
where $\chi_i$ corresponds to isospin $(i-1)/2$, by an orbifold
projection under which the sectors with integer isospin (1,3,5,7) are
even, while those with half-odd-integer isospin,  (2,4,6), are 
odd \cite{gepwit}.  This operation reverses all the ``R-R'' charges, in a 
way reminiscent of what in Section 2 led from the charged to the 
uncharged type II branes \cite{sen}. It is important to note, however,
that in the corresponding partition function
\be
{\cal T}_{D_5} = |\chi_1|^2 + |\chi_3|^2 +  |\chi_5|^2 +
|\chi_7|^2 + |\chi_4|^2 + \chi_2 \bar{\chi}_6 + \chi_6 \bar{\chi}_2 \ ,
\label{c3}
\ee
{\it all} the last three terms, including the diagonal one, 
come from the twisted 
sector of the orbifold, that defines a WZW model on the SO(3)
group manifold. An untwisted charge is thus
transmuted into a twisted one by this orbifold.

The diagonal $A_6$ model
has three types of branes, three corresponding types of anti-branes
and a self-conjugate brane, with R-R charges identified by their
couplings 
to $\chi_2$, $\chi_4$ and $\chi_6$ in the transverse annulus amplitude
\be
\tilde{\cal A}_6 \sim \sum_a \, \frac{\chi_a}{\sin\left( \frac{\pi a}{8}\right)} \,
\left|\, \sum_b \, 
\sin\left ( \frac{\pi a b}{8} \right)\, n_b  \, \right|^2 \ ,
\label{c5}
\ee
determined by
\be
{\cal A}_{A_6} =  \sum_{a,b,k} \, 
{{{\cal A}^k}_{a}}^{b} \ n_a\, \bar{n}_b \, \chi_k
\ , \label{c6}
\ee
where, for the diagonal model, the Cardy ansatz \cite{cardy} identifies
the ranges of $\{a,b\}$ with that of $k$, and the annulus coefficients
${{\cal A}_{ka}}^b$
with the fusion-rule coefficients ${{\cal N}_{ki}}^j$.
The branes with (complex) multiplicities $n_i$ and $n_{8-i}$ have opposite R-R
couplings and, differently from what we saw in Sen's construction \cite{sen} 
for type II theories, one of them, identified by the
multiplicity $n_4$, is
self-conjugate. This would seem to lead to only {\it four} types of
invariant combinations, but actually new twisted R-R couplings emerge,
that distinguish between two branes with opposite twisted charges. 
This provides a geometric picture for the
splitting in \cite{completeness}, while recovering the correct brane
spectrum, as can be seen 
identifying in (\ref{c6})  the pairs of multiplicities 
$n_i$ and $n_{8-i}$. The resulting annulus amplitude, obtained
after an overall rescaling that accounts for the different tensions
of the resulting branes, 
\ba
{\cal A} &=& \left[ n_1 \bar{n}_1+ n_2 \bar{n}_2 + n_3 \bar{n}_3 + 
\frac{n_4 \bar{n}_4}{2}
\right] (\chi_1 + \chi_7) + 
[ n_1 \bar{n}_2 + n_2 \bar{n}_3 + n_3 \bar{n}_4 + h.c. ] 
(\chi_2 + \chi_6) \nonumber \\
&&+
\left[ n_2 \bar{n}_2 + 2 n_3 \bar{n}_3 + \frac{n_4 \bar{n}_4}{2} + 
(n_1 \bar{n}_3 + n_2 \bar{n}_4  + h.c.)
\right] (\chi_3 + \chi_5) \nonumber \\
&&+
[n_1 \bar{n}_4 + 2 n_2 \bar{n}_3 + n_3 \bar{n}_4 + h.c] \chi_4 
 \ , \label{c7}
\ea
acquires a proper particle interpretation after letting $n_4 \to n_4 +
n_5$, while also adding a new {\it twisted} $\chi_4$
coupling in the transverse amplitude, that contributes
\be
\frac{|n_4-n_5|^2}{2} \left( \chi_1 - \chi_3 + \chi_5
- \chi_7 \right) \label{c8}
\ee
to the direct channel. The end result, 
\ba
{\cal A}_{D_5} &=& \left[ n_1 \bar{n}_1+ n_2 \bar{n}_2 +  n_3 \bar{n}_3
+ n_4 \bar{n}_4+ n_5 \bar{n}_5\right] \chi_1  \nonumber \\
&&+
\left[ n_1 \bar{n}_2 + n_2 \bar{n}_3 + n_3 \bar{n}_4 + 
n_3 \bar{n}_5 + h.c. \right] (\chi_2 + \chi_6) \nonumber \\
&&+
\left[ n_2 \bar{n}_2 + 2 n_3 \bar{n}_3 +
(n_1 \bar{n}_3 + n_2 \bar{n}_4 + n_2 \bar{n}_5 + n_4 \bar{n}_5  + h.c.)
\right] \chi_3 \nonumber \\
&&+
[n_1 \bar{n}_4 + n_1 \bar{n}_5 + 2 n_2 \bar{n}_3 + n_3 \bar{n}_4 + 
n_3 \bar{n}_5 + h.c] \chi_4 \nonumber \\
&&+
\left[ n_2 \bar{n}_2 + 2 n_3 \bar{n}_3  + n_4 \bar{n}_4 + n_5 \bar{n}_5
+(n_1 \bar{n}_3 + n_2 \bar{n}_4 + n_2 \bar{n}_5  + h.c.)
\right] \chi_5 \nonumber \\
&&+ \left[ n_1 \bar{n}_1+ n_2 \bar{n}_2 + n_3 \bar{n}_3
+( n_4 \bar{n}_5 + h.c.)\right] \chi_7 
 \ , \label{c9}
\ea
should be compared with the annulus amplitude in \cite{completeness}, 
that follows from it after the orientifold restriction to real
charges, after the redefinitions $n_i \to l_i$, 
$l_1 \leftrightarrow l_2$ and $l_3 \leftrightarrow l_5$. 

In conclusion,
while no brane in the original $A_6$ model was charged with respect
to  $\chi_4$, in the orbifold this 
becomes a twisted sector, and a fractional brane charged with respect to it
(together with its antibrane) appears. The same type of
construction applies to all the other $D_{odd}$ models, and provides
a handy geometric derivation of their brane spectra. This adds to the
original derivation in \cite{completeness}, obtained starting from
the polynomial equations for the boundary couplings extracted from
\cite{cl}, and to the one 
in \cite{bppz}, obtained starting from the completeness 
conditions of \cite{completeness},
\be
{{\cal A}_{ia}}^b \ {\cal A}_{j b c} = \sum_k \ {{\cal N}_{ij}}^k \ 
{\cal A}_{k}^{ac} \ .
\ee
These state that the annulus coefficients satisfy the
fusion algebra or, in equivalent geometrical terms, indeed imply
the completeness of the given brane spectrum.
\vskip 24pt
\begin{flushleft}
{\large \bf Note Added}
\end{flushleft}
The argument relating the $D_{\rm odd}$ charge structure of
\cite{completeness} to the 
emergence of twisted brane couplings was actually anticipated in 
a previous work of Felder, Frohlich, Fuchs and Schweigert \cite{fffs}.
We are grateful to the authors for calling their work to our attention.

\vskip 24pt
\begin{flushleft}
{\large \bf Acknowledgments}
\end{flushleft}
We are grateful to C. Angelantonj, C. Bachas, B. Gato-Rivera, 
L. Huiszoon, A.N. Schelle\-kens,
Ya.S. Stanev and N. Sousa for interesting discussions and comments.
This research was supported in part by the EEC contract 
HPRN-CT-2000-00122, in part by the EEC contract HPRN-CT-2000-00148, and 
in part by the INTAS contract 99-1-590. A.S. would like to thank LPT-Orsay 
and NIKHEF for the kind hospitality during the course of this work and
at its concluding stages and CNRS for financial support.

\appendix
\section{Notation and conventions}

It is often convenient to express the partition 
functions in the NS and R sectors of string models in terms of the
${\rm SO}(p)$ characters ($p$ even)
\ba 
O_{p} &=& {1 \over 2 \eta^{p/2}} 
( \vartheta_3^{p/2} + \vartheta_4^{p/2}) \ , \qquad\quad 
V_{p}={1 \over 2 \eta^{p/2}} ( \vartheta_3^{p/2} - \vartheta_4^{p/2}) \ , 
\nonumber \\ S_{p} &=& {1 \over 2 \eta^{p/2}} ( \vartheta_2^{p/2} +
i^{p/2} \vartheta_1^{p/2}) \ , 
\qquad C_{p}={1 \over 2 \eta^{p/2}} 
( \vartheta_2^{p/2} - i^{p/2} \vartheta_1^{p/2}) \ ,
\label{a1}
\ea 
where the $\vartheta_i$ are the four Jacobi 
theta-functions with half-integer
characteristics. These characters make the space-time interpretation
of the amplitudes rather transparent, since $O_{p}$ begins with 
a scalar, $V_{p}$ with a vector and $S_{p}$ and $C_{p}$ with the 
two spinors of opposite chiralities. Their decompositions with 
respect to lower-dimensional even orthogonal groups,
\ba
O_8 &=& O_{p-1} O_{9-p} + V_{p-1} V_{9-p} \ , \quad  
V_8 = V_{p-1} O_{9-p} + O_{p-1} V_{9-p} \ , \nonumber \\
S_8 &=& S_{p-1} S_{9-p} + C_{p-1} C_{9-p} \ , \quad
C_8 = S_{p-1} C_{9-p} + C_{p-1} S_{9-p} \, \label{a11}
\ea
reflect the simple class properties of the corresponding Lie algebras.
On the other hand, 
for D$p$ branes with odd-dimensional world volumes one needs $O_p$, $V_p$ 
and the single spinor character
\be
S'_{p} = {1 \over \sqrt{2}} \left({\vartheta_2 \over \eta}\right)^{p/2} \ , 
\label{a01}  
\ee
that involves a rescaling, needed
to give the ground state its proper degeneracy. 
Aside from the torus ${\cal T}$ amplitude, the spectra of 
orientifold models \cite{descendants} require 
the Klein-bottle amplitude ${\cal K}$, the annulus
amplitude ${\cal A}$ and the M\"obius amplitude ${\cal M}$,
whose direct-channel modular parameters are as follows:
\be
 {\rm Klein}: \tau = 2 i \tau_2 \ , \quad
 {\rm Annulus}: \tau = {i t \over 2} \ , \quad
 {\rm Moebius}: \tau = {i t \over 2} + {1 \over 2} 
 \ . \label{a02}
\ee
The first two amplitudes are related to the corresponding 
closed-channel amplitudes
$\tilde{\cal K}$ and $\tilde{\cal A}$ by an $S$ transformation
(corresponding to the redefinition $\tau \to - 1/\tau$ of the modular
parameter), while the third is related to the closed-channel amplitude
$\tilde{\cal M}$ by a $P$ transformation (corresponding to the 
redefinition $i t/2 + 1/2 \to i/2 t + 1/2$ of the modular
parameter). In a CFT with central charge $c$, the action of $T$ 
(corresponding to the redefinition $\tau \to \tau + 1$ of the
modular parameter) on a generic basis of characters $\chi_i$
with conformal weights $h_i$ is described by a diagonal unitary matrix
\be
T_{ij} = e^{2 \pi i (h_i - c/24)} \, \delta_{ij} \ ,
\ee
while, when all distinct sectors are described by different 
characters, $S$ and $P$ act as symmetric unitary matrices.

In writing the M\"obius amplitude, for which $\tau$ is not purely imaginary,
it is convenient to work with the real basis of characters
\be
\hat \chi_r = e^{-i \pi h_r} \chi_r \ . \label{a021} 
\ee
While on the $\{\chi_i\}$ $P$ would act as the sequence $T S T^2 S$,  
on the real $\{\hat{\chi}_i\}$ basis
\be
P = T^{1/2} S T^2 S T^{1/2} \ ,
\ee
where $T^{1/2}$ denotes a diagonal unitary matrix with eigenvalues
$e^{i \pi (h_i - c/24)}$.
The effect of these transformations on the characters $O_p$, $V_p$, $S_p$ and
$C_p$ can be deduced from the corresponding transformations of the
Jacobi 
theta-functions. For {\it even} $p$ the $S$ and $P$ matrices are thus \cite{bianchias}
\be  
S_p =  {1 \over 2}
\left(
\begin{array}{cccc} 1 & 1 & 1 & 1 
\\ 1 & 1 & -1 & -1 \\ 1 & -1 & e^{-i p \pi \over 4} & -e^{-i p \pi \over
4} \\ 1 & -1 & -e^{-i p \pi \over 4} & e^{-i p \pi \over 4}
\end{array}
\right) \ , \  P_p = 
\left(
\begin{array}{cccc}  
c & s & 0 & 0 
\\  s & -c & 0 & 0 \\ 0 & 0 & \zeta c & i \zeta s \\
0 & 0 & i
\zeta s & \zeta c
\end{array}
\right) \ , \label{a2}
\ee   
where $c= \cos ({p \pi /8})$, $s= \sin ({p \pi /8})$ 
and $\zeta= e^{-i{p \pi/8}}$ \cite{bianchias}, and 
where a phase ambiguity is fixed uniquely demanding that
\be
S^2 = (S T)^3 \ .
\ee
For odd $p$, 
when a single spinor class is present, the spinorial character
is to be rescaled as in (\ref{a01}). The three resulting
characters have the Ising-like $S$ and $P$ matrices
\be
S_p =  {1 \over 2}
\left(
\begin{array}{cccc} 1 & 1 & \sqrt{2} 
\\ 1 & 1 & -\sqrt{2} \\ \sqrt{2} & -\sqrt{2} & 0 \end{array}
\right) \ , \quad
P_p = 
\left(
\begin{array}{cccc} c & s & 0 
\\ s & -c & 0 \\ 0 & 0 & 1 \end{array}
\right) \ ,
\ee
where $c=\cos(\pi p/8)$ and $s=\sin(\pi p/8)$. 

Notice that, since all these $S$ and $P$ matrices have an analytic dependence on $p$,
one can also formally continue them to zero or
negative values. This suffices to describe the charge assignments
for D0 and ${\rm D}(-1)$
branes, without the need to revert to the covariant formulation.

All the Chan-Paton assignments described in the text rest on the peculiar
fusion rules of the space-time characters \cite{bert}, 
for which the vector class $V_{p-1}$
plays the role of the identity. For instance, for $p-1=4k$, the
two spinor classes are self-conjugate, and these would read
\ba
&& [S_{p-1}]\, [S_{p-1}] = [C_{p-1}]\, [C_{p-1}] =  [O_{p-1}]\, [O_{p-1}] =  
[V_{p-1}]\, [V_{p-1}] = [V_{p-1}] \ , \nonumber \\  
&& [S_{p-1}]\, [C_{p-1}] = [O_{p-1}]\, [V_{p-1}] =  [O_{p-1}] \ ,
\ea
to be compared with the standard fusion rules for the internal
characters, that would read
\ba
&& [S_{9-p}]\, [S_{9-p}] = [C_{9-p}]\, [C_{9-p}] =  [O_{9-p}]\, [O_{9-p}] =  
[V_{9-p}]\, [V_{9-p}] = [O_{9-p}] \ , \nonumber \\ 
&& [S_{9-p}]\, [C_{9-p}] = [O_{9-p}]\, [V_{9-p}] =  [V_{9-p}] \ ,
\ea
so that $V$, rather than $O$, plays the role of identity in the fusion
of space-time characters. For $p-1=4k+2$, similar results hold, but
the two spinor classes are mutually conjugate, so that
\be
[S_{p-1}]\, [S_{p-1}] = [C_{p-1}]\, [C_{p-1}] =  [O_{p-1}] \, \qquad 
[S_{p-1}]\, [C_{p-1}] = [V_{p-1}] 
\ee
and 
\be
[S_{9-p}]\, [S_{9-p}] = [C_{9-p}]\, [C_{9-p}] =  [V_{9-p}] \, \qquad 
[S_{9-p}]\, [C_{9-p}] = [O_{9-p}] 
\ee
for the two cases of space-time and internal characters.
This unusual feature, first elucidated in \cite{bert}, can be related to
the behavior of world-sheet ghosts, and played a key role in \cite{bianchias} 
and in the following work on open-string spectra. The odd-dimensional
Ising-like fusion
\be
[S'_{p-1}]\, [S'_{p-1}] = [O_{p-1}] + [V_{p-1}] 
\ee
also plays a role for some of the branes with odd-dimensional world volumes.

For the sake of brevity, all amplitudes are presented in the text omitting
modular integrals, contributions of space-time bosons and
some overall factors that reflect the brane tensions. 
Thus, for D$p$ branes, with $p+1$ longitudinal dimensions and
$9-p$ transverse dimensions, the complete string amplitudes would 
be
\ba
\!\!\!& &{1 \over (4\pi^2 \alpha')^{5}} 
\int {d^2 \tau \over \tau_2^{6}} 
{  {\cal T} \over |\eta (\tau)|^{16}} \ , \quad
{1 \over (4\pi^2 \alpha')^{5}} 
\int {d\tau_2 \over {\tau_2}^{6}}  { {\cal K} \over {\eta (2 i
\tau_2)}^{8}} \ , \nonumber \\
\!\!\!& & {1 \over (8 \pi^2 \alpha')^{(p+1)/2}}  
\int {dt \over t^{(p+3)/2}} { {\cal A}_{pp} \over {\eta (i t/2)}^{8}}
\ , \nonumber \\
&& {1 \over (8 \pi^2 \alpha')^{(p+1)/2}}
\int {dt \over t^{(p+3)/2}} 
{ {\cal M}_{p} \over {\hat{\eta} (i t/2 + 1/2)}^{p-1}}
\left(\frac{2 \eta}{\vartheta_2}\right)^{(9-p)/2}  \ ,
 \label{i1}
\ea
while the full additional $p$-$q$ amplitudes ($p > q$) 
are
\be
{1 \over (8 \pi^2 \alpha')^{(q+1)/2}}  
\int {dt \over t^{(q+3)/2}} { {\cal A}_{pq} \over {\eta (i t/2)}^{8-p+q}}
\left(\frac{\eta}{\vartheta_4}\right)^{(p-q)/2} \ .
\label{i111}
\ee
The contributions of space-time fermions and 
internal bosons and fermions in the ``amputated'' amplitudes are
sufficient to encode the corresponding GSO projections.

The one-loop amplitudes have dual interpretations as 
tree-level closed-string exchanges between boundaries and crosscaps,
or equivalently between D-branes and O-planes. 
The corresponding closed-string modulus, $l$, is related to the parameters 
in (\ref{a02}) by
\be
 {\rm Klein}: l = {1 \over 2 \tau_2} \ , \quad 
 {\rm Annulus}: l = {2 \over t}  \ , \quad
 {\rm Moebius}: l = {1 \over 2t} \ . 
\label{i2}
\ee 
In the closed-string channel, the complete amplitudes (\ref{i1}) would thus become
\ba
&& {1 \over (4\pi^2 \alpha')^{5}} 
\int dl \ {{\tilde {\cal K}} \over {\eta (i l)}^{8}}  \ , \qquad
{1 \over (8 \pi^2 \alpha')^{(p+1)/2}}  
\int dl \ {{\tilde {\cal A}_{pp}} \over {l^{(9-p)/2} \, \eta (il)}^{8}}
\ ,  \nonumber \\ 
&&{1 \over (8 \pi^2 \alpha')^{(p+1)/2}}
\int dl \ {{\tilde {\cal M}_{p}} \over {\hat{\eta} (il\!+\! 1/2)}^{p-1}} 
\left(\frac{2
\eta}{\vartheta_2}\right)^{(9-p)/2} \ , \nonumber \\
&&{1 \over (8 \pi^2 \alpha')^{(q+1)/2}}  
\int {dl \over l^{(9-p)/2}} \ { \tilde{\cal A}_{pq} \over {\eta (il)}^{8-p+q}}
\left(\frac{2 \eta}{\vartheta_2}\right)^{(p-q)/2} \ , 
\label{i3}
\ea
where, as we have seen, ${\tilde {\cal K}}$, ${\tilde {\cal A}}$ 
(${\tilde {\cal M}}$) are related to the loop amplitudes ${\cal K}$,
${\cal A}$ (${\cal M}$) by the $S$ ($P$) transformation defined in (\ref{a2}).

Notice that, with lower-dimensional branes, some transverse
amplitudes contain powers of $l$, that signal momentum flow across
them. As a result, their factorization properties are
slightly obscured, but they are nonetheless neatly revealed if these
powers are traded for corresponding momentum integrals, 
exposing the propagation of the individual closed-string excitations.

\vskip 28pt



\begin{thebibliography}{99}

\bibitem{dbranes}
J.~Dai, R.~G.~Leigh and J.~Polchinski,
Mod.\ Phys.\ Lett.\ A {\bf 4} (1989) 2073;
P.~Horava,
Nucl.\ Phys.\ B {\bf 327} (1989) 461;
Phys.\ Lett.\ B {\bf 231} (1989) 251;
M.~B.~Green,
Phys.\ Lett.\ B {\bf 329} (1994) 435
[hep-th/9403040].

\bibitem{polchinski} J.~Polchinski,
Phys.\ Rev.\ Lett.\  {\bf 75} (1995) 4724
[hep-th/9510017];
C.~Bachas,
Phys.\ Lett.\ B {\bf 374} (1996) 37
[hep-th/9511043].

\bibitem{descendants}
A.~Sagnotti,
ROM2F-87-25
{\it Talk presented at the Cargese Summer Institute on Non-Perturbative
Methods in Field Theory, Cargese, France, Jul 16-30, 1987};
G.~Pradisi and A.~Sagnotti,
Phys.\ Lett.\ B {\bf 216} (1989) 59;
M.~Bianchi and A.~Sagnotti,
Phys.\ Lett.\ B {\bf 247} (1990) 517;
Nucl.\ Phys.\ B {\bf 361} (1991) 519;
M.~Bianchi, G.~Pradisi and A.~Sagnotti,
Nucl.\ Phys.\ B {\bf 376} (1992) 365.
For a review see:
E.~Dudas,
Class.\ Quant.\ Grav.\  {\bf 17} (2000) R41
[hep-ph/0006190].

\bibitem{greenschwarz} 
M.~B.~Green and J.~H.~Schwarz,
Phys.\ Lett.\ B {\bf 149} (1984) 117;
Phys.\ Lett.\ B {\bf 151} (1985) 21.

\bibitem{bianchias}
M. Bianchi and A. Sagnotti, in \cite{descendants}.

\bibitem{as95} A.~Sagnotti,
hep-th/9509080,
Nucl.\ Phys.\ Proc.\ Suppl.\  {\bf 56B} (1997) 332
[hep-th/9702093].

\bibitem{dhsw}
L.~J.~Dixon and J.~A.~Harvey,
Nucl.\ Phys.\ B {\bf 274} (1986) 93;
N.~Seiberg and E.~Witten,
Nucl.\ Phys.\ B {\bf 276} (1986) 272.

\bibitem{cs} 
C.~Lovelace,
Phys.\ Lett.\ B {\bf 34} (1971) 500;
E.~Cremmer and J.~Scherk,
Nucl.\ Phys.\ B {\bf 50} (1972) 222; 
L.~Clavelli and J.~A.~Shapiro,
Nucl.\ Phys.\ B {\bf 57} (1973) 490;
M.~Ademollo, A.~D'Adda, R.~D'Auria, F.~Gliozzi, E.~Napolitano, S.~Sciuto and P.~Di Vecchia,
Nucl.\ Phys.\ B {\bf 94} (1975) 221.

\bibitem{bstates} C.~G.~Callan, C.~Lovelace, C.~R.~Nappi and S.~A.~Yost,
Nucl.\ Phys.\ B {\bf 293} (1987) 83;
J.~Polchinski and Y.~Cai,
Nucl.\ Phys.\ B {\bf 296}, 91 (1988);
N.~Ishibashi,
Mod.\ Phys.\ Lett.\ A {\bf 4} (1989) 251.
For a recent review, see
M.~R.~Gaberdiel,
Class.\ Quant.\ Grav.\  {\bf 17}, 3483 (2000)
[hep-th/0005029].

\bibitem{dbranemore}A (partial) list of D-brane constructions is:
C.~G.~Callan and I.~R.~Klebanov,
Nucl.\ Phys.\ B {\bf 465} (1996) 473
[hep-th/9511173];
M.~Li,
Phys.\ Rev.\ D {\bf 54} (1996) 1644
[hep-th/9512042];
E.~G.~Gimon and J.~Polchinski,
Phys.\ Rev.\ D {\bf 54} (1996) 1667
[hep-th/9601038];
E.~G.~Gimon and C.~V.~Johnson,
Nucl.\ Phys.\ B {\bf 477} (1996) 715
[hep-th/9604129];
A.~Dabholkar and J.~Park,
Nucl.\ Phys.\ B {\bf 477} (1996) 701
[hep-th/9604178];
F.~Hussain, R.~Iengo, C.~Nunez and C.~A.~Scrucca,
Phys.\ Lett.\ B {\bf 409} (1997) 101
[hep-th/9706186];
P.~Di Vecchia, M.~Frau, I.~Pesando, S.~Sciuto, A.~Lerda and R.~Russo,
Nucl.\ Phys.\ B {\bf 507} (1997) 259
[hep-th/9707068];
M.~Srednicki,
JHEP {\bf 9808} (1998) 005
[hep-th/9807138];
B.~Craps and F.~Roose,
Phys.\ Lett.\ B {\bf 445} (1998) 150
[hep-th/9808074];
E.~Eyras and S.~Panda,
JHEP {\bf 0105} (2001) 056
[hep-th/0009224].
For a review, see:
J.~Polchinski, S.~Chaudhuri and C.~V.~Johnson,
hep-th/9602052;
C.~P.~Bachas,
hep-th/9806199;
C.~V.~Johnson,
hep-th/0007170.


\bibitem{lerda}
M.~Frau, L.~Gallot, A.~Lerda and P.~Strigazzi,
Nucl.\ Phys.\ B {\bf 564} (2000) 60
[hep-th/9903123];
P.~Di Vecchia, M.~Frau, A.~Lerda and A.~Liccardo,
Nucl.\ Phys.\ B {\bf 565} (2000) 397
[hep-th/9906214];
L.~Gallot, A.~Lerda and P.~Strigazzi,
Nucl.\ Phys.\ B {\bf 586} (2000) 206
[hep-th/0001049].

\bibitem{sen} 
A.~Sen,
JHEP {\bf 9806} (1998) 007
[hep-th/9803194],
{\bf 9808} (1998) 010
[hep-th/9805019],
{\bf 9808} (1998) 012
[hep-th/9805170],
O.~Bergman and M.~R.~Gaberdiel,
Phys.\ Lett.\ B {\bf 441} (1998) 133
[hep-th/9806155];
A. Sen,
JHEP {\bf 9809} (1998) 023
[hep-th/9808141],
{\bf 9810} (1998) 021
[hep-th/9809111],
{\bf 9812} (1998) 021
[hep-th/9812031].
For reviews see:\\
A.~Sen,
hep-th/9904207;
A.~Lerda and R.~Russo,
Int.\ J.\ Mod.\ Phys.\ A {\bf 15} (2000) 771
[hep-th/9905006].

\bibitem{ssclosed}
R.~Rohm,
Nucl.\ Phys.\ B {\bf 237} (1984) 553;
S.~Ferrara, C.~Kounnas and M.~Porrati,
Phys.\ Lett.\ B {\bf 181} (1986) 263,
Nucl.\ Phys.\ B {\bf 304} (1988) 500,
Phys.\ Lett.\ B {\bf 206} (1988) 25;
C.~Kounnas and M.~Porrati,
Nucl.\ Phys.\ B {\bf 310} (1988) 355;
S.~Ferrara, C.~Kounnas, M.~Porrati and F.~Zwirner,
Nucl.\ Phys.\ B {\bf 318} (1989) 75;
C.~Kounnas and B.~Rostand,
Nucl.\ Phys.\ B {\bf 341} (1990) 641;
I.~Antoniadis and C.~Kounnas,
Phys.\ Lett.\ B {\bf 261} (1991) 369;
E.~Kiritsis and C.~Kounnas,
Nucl.\ Phys.\ B {\bf 503} (1997) 117
[hep-th/9703059].

\bibitem{carlo}
C.~Angelantonj,
Phys.\ Lett.\ B {\bf 444} (1998) 309
[hep-th/9810214];
R.~Blumenhagen, A.~Font and D.~Lust,
Nucl.\ Phys.\ B {\bf 558} (1999) 159
[hep-th/9904069];
R.~Blumenhagen and A.~Kumar,
Phys.\ Lett.\ B {\bf 464} (1999) 46
[hep-th/9906234];
K.~Forger,
Phys.\ Lett.\ B {\bf 469} (1999) 113
[hep-th/9909010].

\bibitem{ss} 
J.~Scherk and J.~H.~Schwarz,
Nucl.\ Phys.\ B {\bf 153} (1979) 61.

\bibitem{dienes} J.~D.~Blum and K.~R.~Dienes,
Nucl.\ Phys.\ B {\bf 516} (1998) 83
[hep-th/9707160];
B {\bf 520} (1998) 93
[hep-th/9708016].

\bibitem{ssopen}
I.~Antoniadis, E.~Dudas and A.~Sagnotti,
Nucl.\ Phys.\ B {\bf 544} (1999) 469
[hep-th/9807011];
I.~Antoniadis, G.~D'Appollonio, E.~Dudas and A.~Sagnotti,
Nucl.\ Phys.\ B {\bf 553} (1999) 133
[hep-th/9812118];
B {\bf 565} (2000) 123
[hep-th/9907184];
A.~L.~Cotrone,
Mod.\ Phys.\ Lett.\ A {\bf 14} (1999) 2487
[arXiv:hep-th/9909116];
R.~Blumenhagen and L.~Gorlich,
Nucl.\ Phys.\ B {\bf 551} (1999) 601
[hep-th/9812158];
C.~Angelantonj, I.~Antoniadis and K.~Forger,
Nucl.\ Phys.\ B {\bf 555} (1999) 116
[hep-th/9904092].

\bibitem{ft}
E.~S.~Fradkin and A.~A.~Tseytlin,
Phys.\ Lett.\ B {\bf 158} (1985) 316;
A.~Abouelsaood, C.~G.~Callan, C.~R.~Nappi and S.~A.~Yost,
Nucl.\ Phys.\ B {\bf 280} (1987) 599.

\bibitem{bachas} C.~Bachas,
hep-th/9503030.

\bibitem{luest}
R.~Blumenhagen, L.~Gorlich, B.~Kors and D.~Lust,
Nucl.\ Phys.\ B {\bf 582} (2000) 44
[hep-th/0003024],
JHEP {\bf 0010} (2000) 006
[hep-th/0007024];
R.~Blumenhagen, B.~Kors and D.~Lust,
JHEP {\bf 0102} (2001) 030
[hep-th/0012156];
L.~E.~Ibanez, F.~Marchesano and R.~Rabadan,
hep-th/0105155.

\bibitem{douglas}
M.~Berkooz, M.~R.~Douglas and R.~G.~Leigh,
Nucl.\ Phys.\ B {\bf 480} (1996) 265
[hep-th/9606139];
V.~Balasubramanian and R.~G.~Leigh,
Phys.\ Rev.\ D {\bf 55} (1997) 6415
[hep-th/9611165].

\bibitem{aads}
C.~Angelantonj, I.~Antoniadis, E.~Dudas and A.~Sagnotti,
Phys.\ Lett.\ B {\bf 489} (2000) 223
[hep-th/0007090];
C.~Angelantonj and A.~Sagnotti,
hep-th/0010279.

\bibitem{sugimoto}
S.~Sugimoto,
Prog.\ Theor.\ Phys.\  {\bf 102} (1999) 685
[hep-th/9905159].

\bibitem{bsb}
I.~Antoniadis, E.~Dudas and A.~Sagnotti,
Phys.\ Lett.\ B {\bf 464} (1999) 38
[hep-th/9908023];
C.~Angelantonj,
Nucl.\ Phys.\ B {\bf 566} (2000) 126
[hep-th/9908064];
G.~Aldazabal and A.~M.~Uranga,
JHEP {\bf 9910} (1999) 024
[hep-th/9908072];
G.~Aldazabal, L.~E.~Ibanez and F.~Quevedo,
JHEP {\bf 0001} (2000) 031
[hep-th/9909172];
C.~Angelantonj, I.~Antoniadis, G.~D'Appollonio, E.~Dudas and A.~Sagnotti,
Nucl.\ Phys.\ B {\bf 572} (2000) 36
[hep-th/9911081];
M.~Bianchi, J.~F.~Morales and G.~Pradisi,
Nucl.\ Phys.\ B {\bf 573} (2000) 314
[hep-th/9910228];
C.~Angelantonj, R.~Blumenhagen and M.~R.~Gaberdiel,
Nucl.\ Phys.\ B {\bf 589} (2000) 545
[hep-th/0006033].

\bibitem{dm3} E.~Dudas and J.~Mourad,
Phys.\ Lett.\ B {\bf 514}, 173 (2001)
[arXiv:hep-th/0012071];
G.~Pradisi and F.~Riccioni,
arXiv:hep-th/0107090.

\bibitem{samwess} D.~V.~Volkov and V.~P.~Akulov,
Phys.\ Lett.\ B {\bf 46} (1973) 109;
S.~Samuel and J.~Wess,
Nucl.\ Phys.\ B {\bf 221} (1983) 153,
B {\bf 226} (1983) 289,
B {\bf 233} (1984) 488.

\bibitem{dm1} E.~Dudas and J.~Mourad,
Nucl.\ Phys.\ B {\bf 598}, 189 (2001)
[hep-th/0010179].

\bibitem{reckschom}
A.~Recknagel and V.~Schomerus,
Nucl.\ Phys.\ B {\bf 531} (1998) 185
[hep-th/9712186],
Nucl.\ Phys.\ B {\bf 545} (1999) 233
[hep-th/9811237].

\bibitem{ggs}
A.~Sagnotti,
Phys.\ Lett.\ B {\bf 294}, 196 (1992)
[hep-th/9210127].

\bibitem{completeness}
G.~Pradisi, A.~Sagnotti and Y.~S.~Stanev,
B {\bf 381} (1996) 97
[hep-th/9603097];
J.~Fuchs and C.~Schweigert,
Phys.\ Lett.\ B {\bf 414} (1997) 251
[hep-th/9708141].
For a review see:
A.~Sagnotti and Y.~S.~Stanev,
Fortsch.\ Phys.\  {\bf 44} (1996) 585
[Nucl.\ Phys.\ Proc.\ Suppl.\  {\bf 55B} (1996) 200]
[hep-th/9605042];
G.~Pradisi,
Nuovo Cim.\ B {\bf 112} (1997) 467
[hep-th/9603104].

\bibitem{fss} 
J.~Fuchs and C.~Schweigert,
Nucl.\ Phys.\ B {\bf 558} (1999) 419
[hep-th/9902132],
B {\bf 568} (2000) 543
[hep-th/9908025],
Phys.\ Lett.\ B {\bf 490} (2000) 163
[hep-th/0006181].

\bibitem{cardy}
J.~L.~Cardy,
Nucl.\ Phys.\ B {\bf 324} (1989) 581.

\bibitem{pss}
D.~Fioravanti, G.~Pradisi and A.~Sagnotti,
Phys.\ Lett.\ B {\bf 321} (1994) 349
[hep-th/9311183];
G.~Pradisi, A.~Sagnotti and Y.~S.~Stanev,
Phys.\ Lett.\ B {\bf 354} (1995) 279
[hep-th/9503207],
B {\bf 356} (1995) 230
[hep-th/9506014];
L.~R.~Huiszoon, A.~N.~Schellekens and N.~Sousa,
Phys.\ Lett.\ B {\bf 470} (1999) 95
[hep-th/9909114];
J.~Fuchs, L.~R.~Huiszoon, A.~N.~Schellekens, C.~Schweigert and J.~Walcher,
Phys.\ Lett.\ B {\bf 495} (2000) 427
[hep-th/0007174].

\bibitem{klebanov} 
I.~R.~Klebanov and A.~A.~Tseytlin,
Nucl.\ Phys.\ B {\bf 546} (1999) 155
[hep-th/9811035],
JHEP {\bf 9903} (1999) 015
[hep-th/9901101];
J.~A.~Minahan,
JHEP {\bf 9901} (1999) 020
[hep-th/9811156];
A.~Armoni and B.~Kol,
JHEP {\bf 9907} (1999) 011
[hep-th/9906081];
I.~R.~Klebanov, N.~A.~Nekrasov and S.~L.~Shatashvili,
Nucl.\ Phys.\ B {\bf 591} (2000) 26
[hep-th/9909109];
C.~Angelantonj and A.~Armoni,
Nucl.\ Phys.\ B {\bf 578} (2000) 239
[hep-th/9912257],
Phys.\ Lett.\ B {\bf 482} (2000) 329
[hep-th/0003050];
M.~Bianchi and J.~F.~Morales,
JHEP {\bf 0008} (2000) 035
[hep-th/0006176].

\bibitem{bg}
O.~Bergman and M.~R.~Gaberdiel,
Nucl.\ Phys.\ B {\bf 499} (1997) 183
[hep-th/9701137].

\bibitem{witten} E.~Witten,
JHEP {\bf 9812} (1998) 019
[hep-th/9810188].

\bibitem{ghm} M.~B.~Green, J.~A.~Harvey and G.~Moore,
Class.\ Quant.\ Grav.\  {\bf 14} (1997) 47
[hep-th/9605033];
J.~Mourad,
Nucl.\ Phys.\ B {\bf 512} (1998) 199
[hep-th/9709012];
J.~F.~Morales, C.~A.~Scrucca and M.~Serone,
Nucl.\ Phys.\ B {\bf 552} (1999) 291
[hep-th/9812071];
B.~J.~Stefanski,
Nucl.\ Phys.\ B {\bf 548} (1999) 275
[hep-th/9812088].

\bibitem{strominger}
A.~Strominger,
Phys.\ Lett.\ B {\bf 383} (1996) 44
[hep-th/9512059].

\bibitem{bott} R. Bott and L.W. Tu, {\it Differential forms in Algebraic
Topology} (Springer-Verlag, 1982).

\bibitem{dm2}
E.~Dudas and J.~Mourad,
Phys.\ Lett.\ B {\bf 486} (2000) 172
[hep-th/0004165];
R.~Blumenhagen and A.~Font,
Nucl.\ Phys.\ B {\bf 599} (2001) 241
[hep-th/0011269].

\bibitem{sw} J.~H.~Schwarz and E.~Witten,
JHEP {\bf 0103} (2001) 032
[hep-th/0103099].

\bibitem{fs} W.~Fischler and L.~Susskind,
Phys.\ Lett.\ B {\bf 171} (1986) 383,
B {\bf 173} (1986) 262.

\bibitem{englert}
A.~Casher, F.~Englert, H.~Nicolai and A.~Taormina,
Phys.\ Lett.\ B {\bf 162} (1985) 121;
P.~G.~Freund,
Phys.\ Lett.\ B {\bf 151} (1985) 387;
F.~Englert, H.~Nicolai and A.~Schellekens,
Nucl.\ Phys.\ B {\bf 274} (1986) 315;
F.~Englert, L.~Houart and A.~Taormina,
hep-th/0106235.

\bibitem{maldacena} 
J.~Maldacena,
Adv.\ Theor.\ Math.\ Phys.\  {\bf 2} (1998) 231
[Int.\ J.\ Theor.\ Phys.\  {\bf 38} (1998) 1113]
[hep-th/9711200].

\bibitem{bmo} 
B.~Zhou and C.~Zhu,
hep-th/9905146;
P.~Brax, G.~Mandal and Y.~Oz,
Phys.\ Rev.\ D {\bf 63} (2001) 064008
[hep-th/0005242];
G.~L.~Alberghi, E.~Caceres, K.~Goldstein and D.~A.~Lowe,
arXiv:hep-th/0105205.

\bibitem{thompson} D.~M.~Thompson,
hep-th/0105314.

\bibitem{bert}
F.~Englert, H.~Nicolai and A.~Schellekens,
Nucl.\ Phys.\ B {\bf 274} (1986) 315;
W.~Lerche, D.~Lust and A.~N.~Schellekens,
Nucl.\ Phys.\ B {\bf 287} (1987) 477;
W.~Lerche, A.~N.~Schellekens and N.~P.~Warner,
Phys.\ Rept.\  {\bf 177} (1989) 1.

\bibitem{kakt}
Z.~Kakushadze, G.~Shiu and S.~H.~Tye,
Phys.\ Rev.\ D {\bf 58} (1998) 086001
[hep-th/9803141];
C. Angelantonj, in ref. \cite{bsb}.

\bibitem{z3}
E. Gimon and C.V. Johnson, in \cite{dbranemore};
A. Dabholkar and J. Park, in \cite{dbranemore};
C.~Angelantonj, M.~Bianchi, G.~Pradisi, A.~Sagnotti and Y.~S.~Stanev,
Phys.\ Lett.\ B {\bf 385} (1996) 96
[hep-th/9606169],
B {\bf 387} (1996) 743
[hep-th/9607229];
Z.~Kakushadze and G.~Shiu,
Phys.\ Rev.\ D {\bf 56} (1997) 3686
[hep-th/9705163].

\bibitem{dkt}
M.~R.~Douglas,
J.\ Geom.\ Phys.\  {\bf 28} (1998) 255
[hep-th/9604198];
I.~R.~Klebanov and A.~A.~Tseytlin,
Nucl.\ Phys.\ B {\bf 578} (2000) 123
[hep-th/0002159].

\bibitem{schom}
A.~Y.~Alekseev and V.~Schomerus,
Phys.\ Rev.\ D {\bf 60} (1999) 061901
[hep-th/9812193];
C.~Bachas, M.~Douglas and C.~Schweigert,
JHEP {\bf 0005} (2000) 048
[hep-th/0003037];
A.~Y.~Alekseev and V.~Schomerus,
hep-th/0007096.

\bibitem{ciz}
A.~Cappelli, C.~Itzykson and J.~B.~Zuber,
Nucl.\ Phys.\ B {\bf 280} (1987) 445,
Commun.\ Math.\ Phys.\  {\bf 113} (1987) 1.

\bibitem{gepwit} 
D.~Gepner and E.~Witten,
Nucl.\ Phys.\ B {\bf 278} (1986) 493.

\bibitem{cl}
J.~L.~Cardy and D.~C.~Lewellen,
Phys.\ Lett.\ B {\bf 259} (1991) 274;
D.~C.~Lewellen,
Nucl.\ Phys.\ B {\bf 372} (1992) 654.

\bibitem{bppz}
R.~E.~Behrend, P.~A.~Pearce, V.~B.~Petkova and J.~Zuber,
Nucl.\ Phys.\ B {\bf 570} (2000) 525
[Nucl.\ Phys.\ B {\bf 579} (2000) 707]
[hep-th/9908036].

\bibitem{fffs} G.~Felder, J.~Frohlich, J.~Fuchs and C.~Schweigert,
J.\ Geom.\ Phys.\  {\bf 34} (2000) 162
[arXiv:hep-th/9909030].

\end{thebibliography}
\end{document}